\documentclass[amsmath,amssymb,12pt, eqsecnum]{revtex4}

\usepackage{hyperref}
\usepackage{multirow}
\usepackage{array}
\usepackage{sidecap}
\usepackage{subfigure}
\usepackage{graphicx}
\usepackage{epstopdf}
\usepackage{dcolumn}
\usepackage{bm}

\newcommand{\half}[0]{\frac{1}{2}}

\newcommand{\ld}[0]{\mathcal{L}}

\newcommand{\dd}[0]{\textrm{d}}
\newcommand{\defn}[0]{\equiv}
\newcommand{\diag}[0]{\textrm{diag}}
\newcommand{\qsubrm}[2]{{#1}_{\scriptsize{\textrm{#2}}}}

\newcommand{\xtype}[0]{\textsf{{X}}-type }
\newcommand{\ytype}[0]{\textsf{Y}-type }

\newcommand{\symmb}[0]{\varrho}

\newcommand{\ci}[0]{\textrm{i}}
\newcolumntype{V}{>{\centering\arraybackslash} m{.4\linewidth} }

\def\be{\begin{equation}}
\def\ee{\end{equation}}
\def\bea{\begin{eqnarray}}
\def\eea{\end{eqnarray}}
\def\bse{\begin{subequations}}
\def\ese{\end{subequations}}

\newcommand{\fref}[1]{{Figure \ref{#1}}}

\let\oldsqrt\sqrt
\def\sqrt{\mathpalette\DHLhksqrt}
\def\DHLhksqrt#1#2{%
\setbox0=\hbox{$#1\oldsqrt{#2\,}$}\dimen0=\ht0
\advance\dimen0-0.2\ht0
\setbox2=\hbox{\vrule height\ht0 depth -\dimen0}%
{\box0\lower0.4pt\box2}}

\begin{document}

\title{\xtype and \ytype junction stability in domain wall networks}
\author{Richard A. Battye}
\email{richard.battye@manchester.ac.uk}
\affiliation{Jodrell Bank Centre for Astrophysics, School of Physics and Astronomy, The University of Manchester, Manchester M13 9PL, U.K}
\author{Jonathan A. Pearson}
\email{jp@jb.man.ac.uk}
\affiliation{Jodrell Bank Centre for Astrophysics, School of Physics and Astronomy, The University of Manchester, Manchester M13 9PL, U.K}
\author{Adam Moss}
\email{adammoss@phas.ubc.ca}
\affiliation{Department of Physics \& Astronomy, University of British Columbia, Vancouver, BC, V6T 1Z1 Canada }
\date{\today}
\begin{abstract}
We develop an analytic formalism that allows one to quantify the stability properties of \xtype and \ytype junctions in domain wall networks in two dimensions. A similar approach might be applicable to more general defect systems involving junctions that appear in a range of physical situations, for example, in the context of F- and D-type strings in string theory. We apply this formalism to a particular field theory, Carter's pentavac model, where the strength of the symmetry breaking is governed by the parameter $|\epsilon|< 1$. We find that for low values of the symmetry breaking parameter \xtype junctions will be stable, whereas for higher values an \xtype junction  will separate into two \ytype junctions. The critical angle separating the two regimes is given by  $\qsubrm{\alpha}{c} = 293^{\circ}\sqrt{|\epsilon|}$ and this is confirmed using simple numerical experiments. We go on to simulate the pentavac model from random initial conditions and  we find that the fraction of \xtype junctions to \ytype junctions is higher for smaller $\epsilon$, although \xtype junctions do not appear to survive to late time. We also find that  for small $\epsilon$  the evolution of the number of domain walls $\qsubrm{N}{dw}$ in Minkowski space does not follow the standard $\propto t^{-1}$ scaling law with the deviation from the standard lore being more pronounced as $\epsilon$ is decreased. The presence of dissipation appears to restore the $t^{-1}$ lore.
\end{abstract}

 \maketitle

\section{Introduction}

Understanding the dynamics of junctions in networks of topological defects is of great interest in a large variety of physical systems. The relevance of junctions in cosmic string networks has recently been motivated by developments in string theory where the interaction of F- and D-strings \cite{1126-6708-2004-06-013} can be modeled by junctions between composite $(p,q)$ superstrings \cite{1126-6708-2005-09-011, PhysRevD.78.023503, PhysRevD.80.125030}. They are also relevant in condensed matter systems \cite{Thouless} and in building field theoretic models of dark energy \cite{bucher_spergel_1999, battye_bucher_spergel_1999, PhysRevD.76.023005}. These dark energy models are postulated to be formed by a frozen network of domain walls,  but it has remained a challenge to build models in which the freezing of the network is natural. The differences between systems which favour \ytype and \xtype junctions have been highlighted, for example \cite{PhysRevD.73.123528,PhysRevD.73.123519, PhysRevD.73.123520, PhysRevD.72.023503}, with the perception being that models with \xtype junctions are more likely to lead to frustration. In addition it has been suggested to introduce a field coupled to the domain wall forming field with an unbroken continuous $U(1)$ symmetry and hence a conserved Noether charge \cite{BattyePearson-kvform, BattyePearson-charge} or a discrete topological charge in an $SU(5) \times \mathbb{Z}_2$ theory \cite{PhysRevD.69.043513}. It has been shown that both such models can possibly lead to a frozen network. 

In this paper we aim to develop an understanding of the stability properties of \xtype junctions in the pentavac field theory that was first proposed by Carter \cite{carter_pent_1, carter_x-stability}. The Lagrangian is that for two scalar fields $\Phi = |\Phi|e^{\ci \phi}, \Psi = |\Psi|e^{\ci \psi}$ which is invariant under global $U(1)\times U(1)$ transformations except for the symmetry breaking term in the potential
\bea
\qsubrm{V}{break} = \epsilon \bigg[\cos\left( 2\phi + \psi\right) + \cos\left(2\psi - \phi\right) \bigg]\,.
\eea
Carter showed that this model prefers \xtype over \ytype  junctions in the limit $\epsilon\rightarrow 0$; in this work we will consider a range of values of $\epsilon\ne 0$.

In recent work Avelino et al \cite{avelino_dw_carter}   studied this model and make the claim that, although \xtype junctions do form, their effect is not sufficient to prevent the relaxation of the system in to a scaling regime where the number of domain walls, $N_{\rm dw}(t)$,  scales like $t^{-1}$. Moreover, they suggest that the \xtype junctions which are seen are as a result of the algorithm \cite{1989ApJ...347..590P} they employ to model the expansion of the Universe. However their simulations probe a limited range, taking $0.05 \leq\epsilon \leq 0.2$.

We will present a perturbative calculation of the energy associated with domains walls in this model which we use determine the stability of \xtype junctions with particular intersection angle $\alpha$. This allows us to determine $\qsubrm{\alpha}{c}$, the angle above which the \xtype junction is stable and below which it splits into two \ytype junctions, as a function of $\epsilon$. We check this result against that found in numerical simulations of the full field theory and find excellent quantitative agreement. This   result suggests that, for sufficiently low $\epsilon$, \xtype junctions will be stable in all but the most extreme circumstances, for example, when junctions which have very small $\alpha$ are generated by the dynamics.  We investigate this using simulations starting with random initial conditions.

The structure of this paper is as follows. We explain what junctions are and present the general model we use to investigate junction stability in section \ref{sec:implementation}. In section \ref{sec:ene_config} we present our perturbative analytical method we use to understand the stability of an \xtype  junction against decay into two \ytype junctions. We present Carter's pentavac model in section \ref{sec:pentavac-pot}, and simulate \xtype junctions within the model. We present results from numerical simulations of Carter's pentavac model from random initial conditions in section \ref{sec:pentavac-scaling} and some concluding remarks in section \ref{ref:conc}.

\section{Basic picture}

\subsection{Walls and junctions}
\label{sec:implementation}
\begin{figure*}[!h]
      \begin{center}
\subfigure[\, \ytype and \xtype junctions]{\includegraphics[scale=0.9]{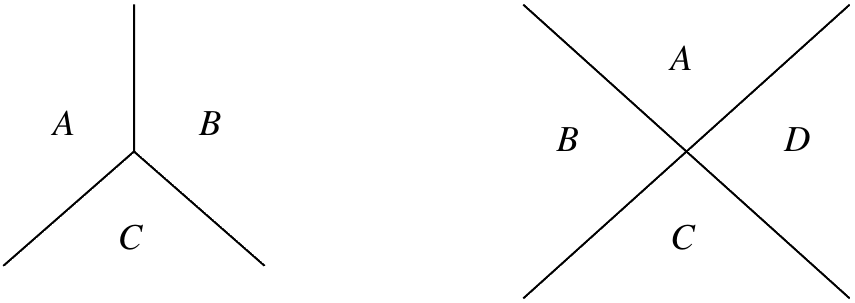}}
\subfigure[\, \xtype junction decay]{\includegraphics[scale=0.5]{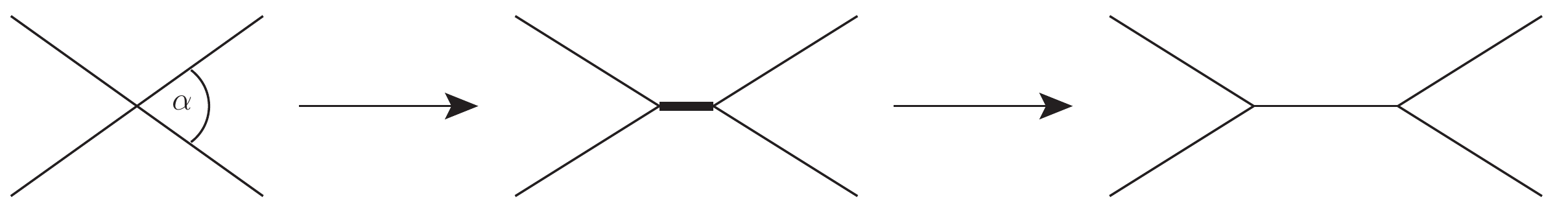}}
\end{center}
\caption{Schematics of  \ytype and \xtype junctions. In (a) we show three vacua $ABC$ cycled around a point to produce \ytype junction and four vacua $ABCD$ an \xtype junction. In (b) we give a schematic view of an \xtype junction decay to two \ytype junctions. The relative thickness of the lines denote the relative tensions of the corresponding domain wall: the thick horizontal wall is a high tension wall $\qsubrm{T}{II}$ and the thin lines are low tension walls $\qsubrm{T}{I}$.}\label{fig:xy}
\end{figure*}

When a field theory has multiple disconnected vacua, the interpolation of the field between spatially adjacent vacuum states is a domain wall (see, for example, \cite{VS}). If more than two vacuum states meet at a point in space the intersection of the set of domain walls produce junctions. Four vacua cycled round a point produce \xtype junctions, and three vauca produce \ytype junctions. See \fref{fig:xy} for a schematic representation of \xtype and \ytype junctions. The tension of the walls is given by the ``distance'' (calculated via non-trivial integration) between points in the field configuration space.

An \xtype junction can decay into two \ytype junctions if the energy of the configuration is reduced, as depicted in \fref{fig:xy}(b). As the \xtype junction splits apart a high tension wall $\qsubrm{T}{II}$ is created, with all other walls being low tension $\qsubrm{T}{I}$. One can think of an \xtype junction as being two \ytype junctions ``glued'' together and separated by a high tension wall. Hence we can investigate their stability by considering whether two \ytype junctions, separated by a high tension wall, are in equilibrium.
 By simple inspection of the middle panel of \fref{fig:xy}(b) and resolving forces,  \xtype junctions are stable if
\begin{eqnarray}
\label{eq:sec2-cond-x-stab-1}
\qsubrm{T}{II} > 2\cos\left(\tfrac{1}{2}\alpha\right) \qsubrm{T}{I}.
\end{eqnarray}
If this condition is satisfied, the horizontal high tension wall is strong enough to overcome repulsion of the two low tension walls, and the \xtype junction retains its form. There will be a critical angle $\qsubrm{\alpha}{c}$, dependent upon the model parameter, above which \xtype junctions will be stable and below which there will be separation into two \ytype junctions.

\subsection{General formalism}
\label{sec:ene_config}
\begin{figure*}[!t]
      \begin{center}
\subfigure[\,Domain wall in spacetime manifold.]{\includegraphics[scale=0.8]{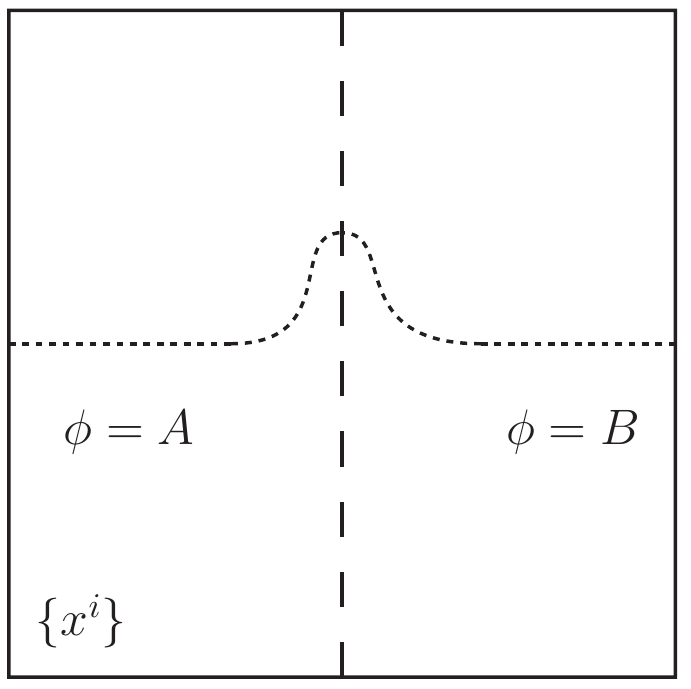}}\qquad \qquad
\subfigure[\, Domain wall trajectory in field manifold. Field values that minimize the potential are displayed by full circles.]{\includegraphics[scale=0.8]{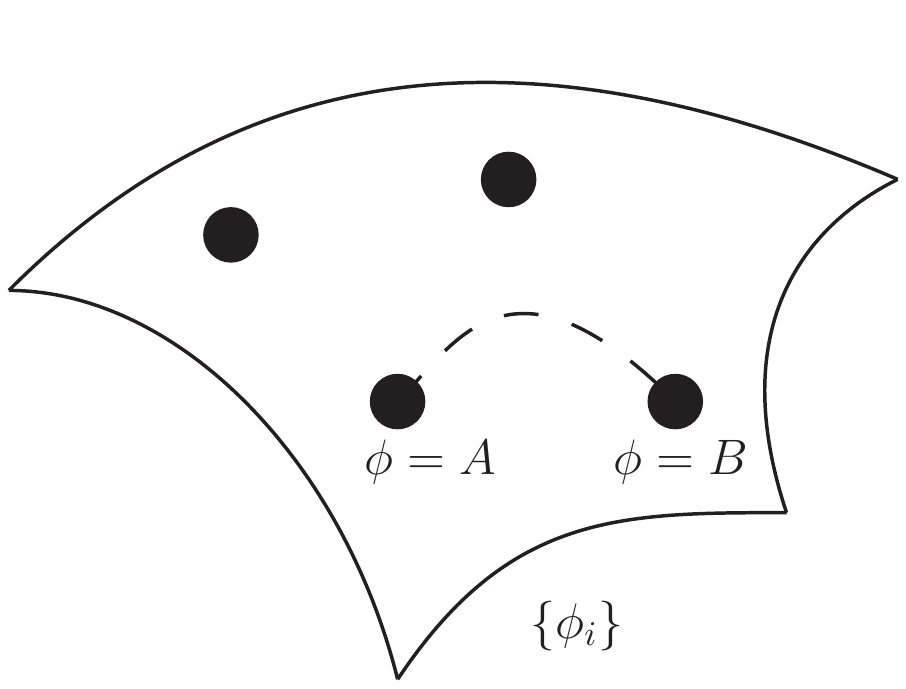}}
      \end{center}
\caption{Schematic of a domain wall configuration. In (a) we show a configuration where there are regions of space occupying different vacua, separated by a domain wall whose energy density is depicted by the dotted line. In (b) we show a field trajectory between two different vacuum states, signifying the presence of a domain wall. }
\label{fig:sec3-dw-schem-field-manf}
\end{figure*}

We will consider a static domain wall configuration, where the field $\phi$ (say) interpolates from $\phi(x = +\infty) = A$ to $\phi(x = -\infty)=B$, giving rise to a domain wall between the vacua $A$ and $B$. Integrating the energy density of the relevant field theory over the domain wall (between $x = \pm\infty$) gives the tension of the wall separating $A$ and $B$. An alternative method of finding the tension is to integrate through the field manifold itself, rather than through the spacetime manifold. We will   show how to convert between these two approaches. In \fref{fig:sec3-dw-schem-field-manf} we schematically depict a domain wall in the spacetime manifold with coordinates $\{x^i\}$ and field manifold with coordinates $\{\phi_i\}$. 

Integrating the energy density $\mathcal{E}$ of a one-dimensional static manifold with Cartesian coordinate $x \in (-\infty, +\infty)$, having Riemannian kinetic metric $G_{ij}$, gives the tension
\begin{eqnarray}
T =E= \int^{\infty}_{-\infty}\dd x \,\mathcal{E} = \int \dd x\,\left( \half G_{ij}\frac{\dd\varphi^i}{\dd x}\frac{\dd \varphi ^j}{\dd x} + V\right),
\end{eqnarray}
where the $\{\varphi^i\}$ are field coordinates and $V$ is a potential function which has zero value at its minimum, $\qsubrm{V}{min} = 0$. Introducing an affine parameter  $\Lambda$ in the field manifold, one finds that the tension can be written as
\begin{eqnarray}
\label{eq:sec2:step-1}
T = \int \dd\Lambda \,\left(\frac{1}{2x'} G_{ij} \varphi'^i \varphi'^j  + Vx' \right),
\end{eqnarray}
where
\bea
x' \defn \frac{\dd x}{\dd \Lambda}, \qquad \varphi' \defn \frac{\dd\varphi}{\dd\Lambda}.
\eea
To eliminate the dependance of the tension on the spatial coordinate $x$, we minimize the integrand with respect to $x'$ (i.e. setting $\delta T/\delta x' = 0$), giving
\begin{eqnarray}
x' = \sqrt{\frac{G_{ij} \varphi'^i \varphi'^j}{2V}},
\end{eqnarray}
which can be substituted into (\ref{eq:sec2:step-1}) to give
\begin{eqnarray}
\label{eq:sec2:step-2}
T = \int \dd \Lambda\,\sqrt{2VG_{ij} \varphi'^i \varphi'^j}.
\end{eqnarray}
This expression   will form the starting point of our subsequent discussions. If the endpoints of integration are taken to be different vacuum states, then $T$ is the tension of the domain wall between the vacua. 

So far the computation of the tension has been performed over some continuous manifold, such as the circle $S^1$ or torus $S^1\times S^1$, having continuous symmetry. The vacuum manifold of a field theory such as the pentavac model is found by constraining motion to some submanifold via a small symmetry breaking parameter $\symmb$, say. We now perturbatively expand the tension (\ref{eq:sec2:step-2}), by writing the potential and metric in orders of $\symmb $, 
\begin{eqnarray}
\label{eq:sec2:step-222}
V = \symmb V^{(1)} + \symmb ^2V^{(2)},\qquad G_{ij} = \delta_{ij} + \symmb G_{ij}^{(1)},
\end{eqnarray}
where $\delta_{ij}$ is the Kronecker-delta symbol and we used the fact that $\qsubrm{V}{min}=0$.   Substituting the expansions (\ref{eq:sec2:step-222}) into (\ref{eq:sec2:step-2}), the tension can be written perturbatively as
\begin{eqnarray}
\label{eq:sec2:step-22}
T\approx \int \dd\Lambda \sqrt{2\symmb V^{(1)}} \left\{1 + \frac{\symmb}{2}\left(\frac{V^{(2)}}{V^{(1)}} +  G_{ij}^{(1)} \varphi'^i \varphi'^j \right) + \mathcal{O}(\symmb^2)\right\},
\end{eqnarray}
where we have used $\delta_{ij}\varphi'^i \varphi'^j=1$.  We require $V^{(1)}\symmb>0$, a necessary condition for this analytic method to work. 
\subsection{Specific form of the energy density}
We now derive an expression that can be used to calculate the tension of a domain wall in models with energy density given by
\begin{eqnarray}
\label{eq:secprop_en_1}
\mathcal{E} = \half \left[ |\nabla\Phi|^2 + |\nabla\Psi|^2 \right] + \frac{\lambda}{4}\left[ \left( |\Phi|^2 - 1\right)^2 + \left( |\Psi|^2 - 1\right)^2\right] + \symmb |\Phi|^2|\Psi|^2F(\phi, \psi),
\end{eqnarray}
where
\begin{eqnarray}
\Phi = |\Phi|e^{\ci\phi},\qquad \Psi = |\Psi|e^{\ci\psi}.
\end{eqnarray}

We now perturb about the solution for the $\symmb=0$ model and integrate along the geodesic trajectories in the vacuum manifold to give the tension. In particular we write
\[
\Phi = \left( 1+\symmb \gamma_1\right)e^{\ci \phi},\qquad \Psi = \left( 1+\symmb \gamma_2\right)e^{\ci \psi}.
\]
Substituting this perturbation into the energy density (\ref{eq:secprop_en_1}) whilst keeping terms up to $\mathcal{O}(\symmb ^2)$ gives
\begin{eqnarray}
\mathcal{E} &= &\half \bigg[ (\nabla\phi)^2 \left(1 + 2\symmb \gamma_1 + \symmb^2 \gamma_1^2 \right) + (\nabla\psi)^2 \left(1 + 2\symmb \gamma_2 + \symmb^2 \gamma_2^2 \right)+ \symmb^2\left[\left(\nabla\gamma_1\right)^2 + \left(\nabla\gamma_2\right)^2\right] \bigg]\nonumber\\
\label{eq:secprop_en_2}
&& +\symmb F(\phi,\psi) + \symmb^2\left[2(\gamma_1 + \gamma_2)F(\phi, \psi) + \lambda(\gamma_1^2 + \gamma_2^2) \right] + \mathcal{O}(\symmb^3).
\end{eqnarray}
This can be written as
\[
\mathcal{E} \approx \half G_{ij} \frac{\dd \varphi^i}{\dd x} \frac{\dd \varphi ^j}{\dd x} + \half M_{ij} \frac{\dd \gamma^i}{\dd x}\frac{\dd \gamma^j}{\dd x}+ V,\qquad \varphi^i = \{\phi, \psi\},\qquad M_{ij} = \symmb^2\delta_{ij},
\]
if one makes the identifications in terms of the expansion (\ref{eq:sec2:step-222})
\[
V^{(1)} = F(\phi, \psi), \qquad V^{(2)} = 2(\gamma_1 + \gamma_2)F(\phi, \psi) + \lambda(\gamma_1^2 + \gamma_2^2),\qquad G_{ij}^{(1)} = 2\,\diag(\gamma_1, \gamma_2).
\]
Then using  (\ref{eq:sec2:step-22}) one finds
\begin{eqnarray}
\label{eq:sec3-ten-funct-1}
T = \int \dd\Lambda \sqrt{2 \symmb F} \left\{ 1+ \symmb \left[\gamma_1\left(1+\phi'^2\right) + \gamma_2\left(1+\psi'^2\right) + \frac{\lambda}{2F}\left(\gamma_1^2 + \gamma_2^2\right) \right] + \mathcal{O}(\symmb^2)\right\}.
\end{eqnarray}
The trajectories $\gamma_i$ that extremise this functional are found to be
\[
\gamma_1 = -\frac{F}{\lambda}\left( 1+\phi'^2\right),\qquad \gamma_2 = -\frac{F}{\lambda}\left( 1+\psi'^2\right),
\]
and, using the condition $\phi'^2 + \psi'^2=1$,  (\ref{eq:sec3-ten-funct-1}) becomes
\begin{eqnarray}
T = \int \dd\Lambda\, \sqrt{2 \symmb F} \left\{ 1 - \symmb\frac{F}{2\lambda} \left( 4+ \phi'^4 + \psi'^4\right) + \mathcal{O}(\symmb^2)\right\}.
\end{eqnarray}
This can be written as
\begin{eqnarray}
T = U - \frac{\symmb}{\lambda} W+ \mathcal{O}(\symmb^2),
\end{eqnarray}
where the zeroth and first order terms in the tension are given by
\begin{eqnarray}
\label{eq:sec2:sugg_form_alphac}
U = \int \dd\Lambda\, \sqrt{2 \symmb}F^{1/2},\qquad W =  A\int \dd\Lambda\, \sqrt{2 \symmb}{F}^{3/2},\qquad A\defn \frac{1}{2} \left( 4+ \phi'^4 + \psi'^4\right).
\end{eqnarray}
To finally obtain stability criteria, one must integrate these expressions for a particular theory (i.e. choice of $F(\phi, \psi)$), over particular trajectories (such as high and low tension walls). This is a task we leave for the next section, where we continue with a specific example of a field theory.

\section{Specific example: the pentavac model}
\label{sec:pentavac-pot}
In this section we will consider a specific field theory: Carter's pentavac model \cite{carter_pent_1, carter_x-stability}. We will begin with a discussion of the salient aspects of the pentavac model and continue to give an understanding of the stability  of \xtype and \ytype junctions. Specifically we will use the perturbative expression for the tension (\ref{eq:sec2:sugg_form_alphac}) to obtain an analytic dependance of the critical internal intersection angle $\qsubrm{\alpha}{c}$, as a function of $\epsilon$.
\subsection{Pentavac potential}
Carter's pentavac model \cite{carter_pent_1,carter_x-stability} has two scalar fields $\Psi(x^{\mu}), \Phi (x^{\mu})\in \mathbb{C}$ whose dynamics are described the Lagrangian density
\begin{eqnarray}
\ld = \half \partial_{\mu}\Phi\partial^{\mu}\bar{\Phi} + \half \partial_{\mu}\Psi \partial^{\mu}\bar{\Psi} - V\left( \Phi, \Psi\right),
\end{eqnarray}
interacting in the potential
\begin{eqnarray}
\label{eq:carterpot-1}
V = \frac{\lambda}{4} \left( |\Phi|^2 -\eta^2\right)^2 +  \frac{\lambda}{4} \left(|\Psi|^2-\eta^2\right)^2 + \symmb |\Phi|^2|\Psi|^2\left(\cos\theta + \cos\chi + V_0\right),
\end{eqnarray}
where
\begin{eqnarray}
\label{eq:Sec4;thetachi-defn}
\Phi = |\Phi|e^{\ci\phi},\quad \Psi = |\Psi|e^{\ci\psi},\qquad \theta \defn 2\phi +\psi,\quad \chi \defn 2\psi-\phi,
\end{eqnarray}
and $V_0$ is chosen so that $\qsubrm{V}{min} = 0$, to be in accord with the requirements of the analytic calculation we will perform.
One can easily show that by rescaling the fields according to $\Phi \mapsto \Phi\eta, \Psi \mapsto \Psi\eta$ and the space-time coordinates $x \mapsto x/\sqrt{\lambda}\eta$, there is only one meaningful combination of model parameters, namely $\epsilon \defn \symmb /\lambda$. Hence, without loss of generality we will set the model parameters $\lambda = \eta =1$ throughout; thus, the limit $\epsilon \rightarrow 0$ corresponds to $\symmb/\lambda \rightarrow 0$ which can be achieved by $\lambda \rightarrow \infty$. With the rescaling the potential (\ref{eq:carterpot-1}) becomes
\begin{eqnarray}
\label{eq:Sec4;thetachi-defn-rescale}
V = \frac{1}{4} \left( |\Phi|^2 -1\right)^2 +  \frac{1}{4} \left(|\Psi|^2-1\right)^2 +{\epsilon} |\Phi|^2|\Psi|^2\left(\cos\theta + \cos\chi + V_0\right).
\end{eqnarray}
So that the $\epsilon<0$ sector can be understood on the same analytical footing as the $\epsilon>0$ calculation presented in the previous section, we must be able to identify a transformation that relates the $\epsilon>0$ and $\epsilon<0$ sectors. For the pentavac model let us write
\bea
\label{eq:sec35ref}
J(\theta,\chi,\epsilon,V_0) \defn \epsilon(\cos\theta+\cos\chi +V_0),
\eea
then, the transformation (and invariance) is
\bea
\label{eq:sec4-pent-trans-iv-1}
J(\theta,\chi,\epsilon,V_0)\longmapsto \tilde{J}(\theta,\chi,\epsilon,V_0)=J(\theta+\pi,\chi+\pi,-\epsilon,-V_0)\equiv J(\theta,\chi,\epsilon,V_0).
\eea

If $\epsilon =0$ the theory has continuous  $U(1)\times U(1)$ symmetry, meaning that the vacuum manifold is a torus, $S^1\times S^1$. When $\epsilon \neq 0$ the continuous symmetry is broken and there are discrete minima on the torus. One can easily find that in the vacuum the field moduli are
\begin{eqnarray}
\label{eq:sec4:vac-mod}
|\Phi|^2 = |\Psi|^2 = \frac{1}{1 - 4|\epsilon|}.
\end{eqnarray}

 This specific choice of symmetry breaking potential is periodic, and therefore it is natural to look for transformations of the phases $(\phi, \psi)$ that leave the Lagrangian invariant. Transformations of the form
\begin{eqnarray}
\phi\longmapsto \tilde{\phi} = \phi + \delta_{\phi},\qquad \psi\longmapsto \tilde{\psi} = \psi + \delta_{\psi},
\end{eqnarray}
leave the symmetry breaking potential unchanged if the argument of the cosine-terms in the potential increment by an integer multiple of $2\pi$, i.e. if
\begin{eqnarray}
2\phi + \psi \longrightarrow 2\phi + \psi + 2\pi n,\qquad 2\psi - \phi \longrightarrow 2\psi - \phi + 2\pi m,
\end{eqnarray}
where $n,m\in \mathbb{Z}$ and hence
\begin{eqnarray}
\label{eq:sec-pent-wind-numbers}
\delta_{\phi} = \frac{2\pi}{5}\left( 2n -m\right),\qquad \delta_{\psi} = \frac{2\pi}{5}\left( 2m + n\right).
\end{eqnarray}
Therefore, we can introduce two winding numbers $(n,m)$ which can be used to construct transformations of the fields which leave the Lagrangian density invariant, and therefore construct the phase combinations that constitute the vacuum manifold. In the  $\epsilon <0$ sector the  five vacua of the theory are given in terms of the phases $(\phi,\psi)$ by
\[
(0,0),\quad A:(-2\pi/5,4\pi/5), \quad B:(4\pi/5, 2\pi/5),\quad  C:(2\pi/5, -4\pi/5), \quad D:(-4\pi/5, -2\pi/5),
\]
and when $\epsilon >0$, by the phase combinations
\[
(\pi,\pi),\quad A:(-3\pi/5,\pi/5)\quad B:(\pi/5,3\pi/5),\quad C:(3\pi/5,-\pi/5),\quad D:(-\pi/5,-3\pi/5).
\]

\begin{figure*}[!t]
      \begin{center}
      {\includegraphics[scale=0.5]{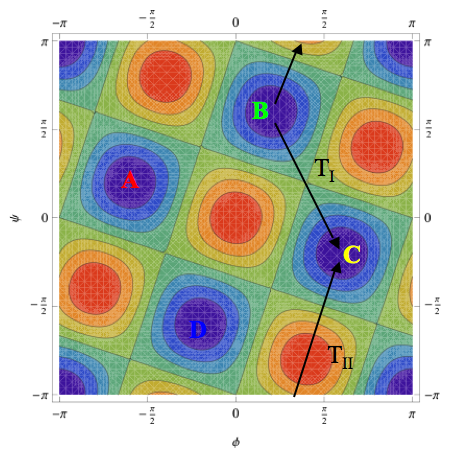}}
      \end{center}
\caption{The vacuum manifold of the pentavac model in the  $\epsilon >0$ sector. We have marked on two wall trajectories between the vacua $B$ and $C$; the low tension interpolation,  $\qsubrm{T}{I}$, and high tension, $\qsubrm{T}{II}$, (remembering that this grid has the topology of a torus). The minima of the theory correspond to the blue (dark) regions, and the maxima to the orange (light) regions.}
\label{penta-vac-schematic-eppos}
\end{figure*}

We present an illustration of the vacuum manifold in \fref{penta-vac-schematic-eppos} in the $\epsilon >0$ sector (the $\epsilon <0$ manifold has the same structure, but with the maxima and minima reversed). The plot we present is in the space of field phases, $(\phi, \psi)$, where the figure is periodic in both directions (i.e. toroidal). Marked onto the figure are two energetically different trajectories between the vacua $B$ and $C$, i.e. the two different types of domain wall: high tension ${\qsubrm{T}{II}}$ and low tension $\qsubrm{T}{I}$. Two such trajectories exist between all pairs of vacua.

In the limit $\epsilon\rightarrow 0$ the vacuum $|\Phi| = |\Psi| \rightarrow 1$, which has the interesting consequence of decoupling the field phases; that is, taking $|\Phi| = |\Psi| =1$ the Lagrangian density can be written to leading order in $\epsilon$ as
\begin{eqnarray}
\ld = \frac{1}{10} (1-4\epsilon)\left[ \partial_{\mu}\theta\partial^{\mu}\theta + \partial_{\mu}\chi \partial^{\mu}\chi\right] - 2\epsilon \left( \cos^2(\theta/2) + \cos^2(\chi/2) - 1\right) + \mathcal{O}(\epsilon^2).
\end{eqnarray}
This describes two non-interacting sine-Gordon kinks and provides us with a very useful result: in the asymptotic $\epsilon\rightarrow 0$ limit an \xtype junction is absolutely stable.  

\subsection{Analytic calculation of $\qsubrm{\alpha}{c}(\epsilon)$}
If we define $f = J/\epsilon$ then for $\epsilon >0$, $f(\phi, \psi) = \cos\theta + \cos\chi +2$ (n.b. for $\epsilon <0$, $f(\phi, \psi) = \cos\theta + \cos\chi - 2$) and we used this to compute (\ref{eq:sec2:sugg_form_alphac}). Now we pick typical high and low tension trajectories and perform the necessary steps to compute $\qsubrm{\alpha}{c}(\epsilon)$:
\begin{itemize}
\item \textit{\textbf{Low tension}} $\qsubrm{T}{I}$: the low tension trajectory has $\chi = \pi$, so that $\qsubrm{f}{I} = 2\cos^2(\theta/2)$ and $\dd\Lambda=\frac{1}{\sqrt{5}}\dd\theta$. The variation of the field phases $\phi, \psi$ along a low tension trajectory is found by inspecting the vacuum manifold and deducing the equation of a line linking vacua. One finds  $\phi'^2 = 1/5, \psi'^2 = 4/5$.
\item \textit{\textbf{High tension}} $\qsubrm{T}{II}$: the high tension has $\chi = \theta$, so that $\qsubrm{f}{II} = 4\cos^2(\theta/2)$ and $\dd\Lambda = \frac{1}{\sqrt{5}}\sqrt{2}\dd\theta$. The variation of the field phases along a high tension trajectory is found to be $\phi'^2 = 1/10, \psi'^2= 9/10$.
\end{itemize}
From these expressions one can find that the zeroth order tensions satisfy
\bea
\qsubrm{U}{II} = 2\qsubrm{U}{I}.
\eea
Using these ingredients in (\ref{eq:sec2:sugg_form_alphac}), one can easily compute
\begin{eqnarray}
\label{eq:sec2:sugg_form_alphac_2}
\frac{\qsubrm{W}{II}}{\qsubrm{U}{II}} = \frac{1928}{300},\qquad \frac{\qsubrm{W}{I}}{\qsubrm{U}{I}} = \frac{468}{150}.
\end{eqnarray}

The stability of an \xtype junction is assured if the internal intersection angle $\alpha$ satisfies (\ref{eq:sec2-cond-x-stab-1}). We take $\qsubrm{\alpha}{c}$ to be the intersection angle at equality. In the notation of (\ref{eq:sec2:sugg_form_alphac}), we can expand to first order to find
\[
\frac{\qsubrm{T}{II}}{2\qsubrm{T}{I}} = \cos\left(\tfrac{1}{2}\qsubrm{\alpha}{c}\right) = \frac{\qsubrm{U}{II}}{2\qsubrm{U}{I}}\left[ 1-\epsilon \left( \frac{\qsubrm{W}{II}}{\qsubrm{U}{II}} - \frac{\qsubrm{W}{I}}{\qsubrm{U}{I}}\right)+ \mathcal{O}(\epsilon^2)\right].
\]
Using our computed values (\ref{eq:sec2:sugg_form_alphac_2}), this becomes
\begin{eqnarray}
\cos\left(\tfrac{1}{2}\qsubrm{\alpha}{c}\right) = 1-{\epsilon}{}\frac{248}{75} +\mathcal{O}(\epsilon^2).
\end{eqnarray}
and expanding $\cos\left(\tfrac{1}{2}\qsubrm{\alpha}{c}\right) \approx 1 - \tfrac{1}{8}\qsubrm{\alpha}{c}^2$, one finds
\bea
\label{eq:sec_ep-alphac-realthips}
\qsubrm{\alpha}{c} \approx 293^{\circ}\sqrt{\epsilon},
\eea
which holds for $0 < \epsilon \ll 1$. In fact, we see that by the transformation/invariance relationship (\ref{eq:sec4-pent-trans-iv-1}) we can extend (\ref{eq:sec_ep-alphac-realthips}) to include $\epsilon <0$, and write $\qsubrm{\alpha}{c} \approx 293^{\circ}\sqrt{|\epsilon|}$.
Hence, we see that the expression predicts that systems with low $\epsilon$ are more stable to acute perturbations that those systems with high $\epsilon$.

\subsection{Numerical simulation of \xtype junctions}
In order to check the veracity of the analytic calculation,  we have performed a suite of simulations evolving the equations of motion. We will investigate the stability of \xtype junctions by constructing an \xtype junction with a particular $\alpha$ and investigate the conditions under which it decays, if indeed it does decay. 

\begin{figure*}[!t]
      \begin{center}
\includegraphics[scale=0.9]{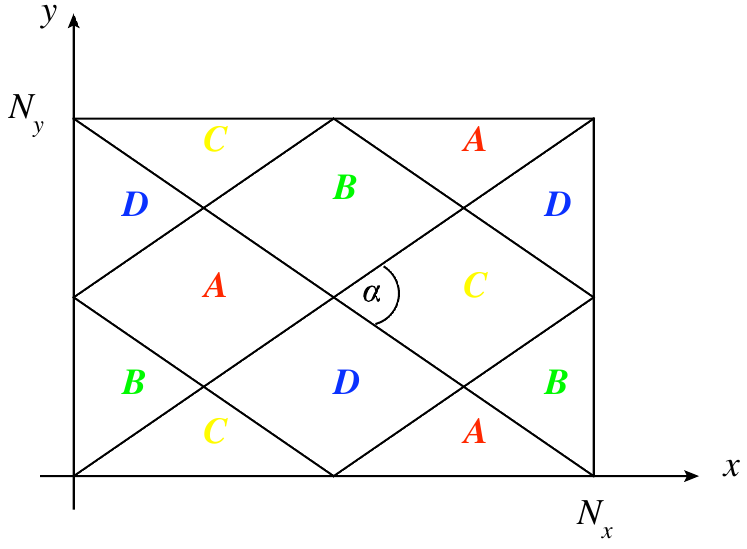}
\end{center}
\caption{The initial field configuration for setting up the \xtype  junction under investigation. There are 8 equivalent \xtype  junctions in this configuration -- with the figure being periodic in both directions.}\label{fig:x-junc-schematic}
\end{figure*}

In our numerical investigations we discretize space to fourth order, and time to second order -- evolution is performed with a leapfrog algorithm, on a grid of $N_x\times N_y$ grid-points. We use space step-size $\Delta x = 0.5$ and time step-size  $\Delta t = 0.1$. To smooth out initial unphysical discontinuities we damp the fields for the first 200 time-steps (i.e. $t < 20$) and then remove damping, allowing the system to freely evolve.  All simulations use periodic boundary conditions. We will display images of the field, where each colour corresponds to the vacua closest to the field at a given location. The initial conditions for our numerical simulations have \xtype junctions with a ``tunable'' initial internal angle $\alpha$ by choosing $\alpha = 2\tan^{-1}\left( N_y/N_x\right)$ as shown in \fref{fig:x-junc-schematic}.

To make the model numerically tractable we cast each complex field as two real fields, $\Phi = \phi_1 + \ci \phi_2$, $\Psi = \phi_3+\ci \phi_4$, and then one finds that the potential can be written as
\begin{eqnarray}
V(\phi_i) &=& \frac{1}{4} \left( \phi_1^2 + \phi_2^2-1\right)^2 +  \frac{1}{4}\left( \phi_3^2 + \phi_4^2 -1\right)^2 \nonumber\\
&&+ \epsilon \left[\sqrt{\phi_3^2 + \phi_4^2} \left( \phi_3(\phi_1^2 - \phi_2^2) - 2\phi_1\phi_2\phi_4\right) +\sqrt{\phi_1^2 + \phi_2^2} \left( \phi_1(\phi_3^2 - \phi_4^2) + 2\phi_2\phi_3\phi_4\right) \right].\nonumber\\
\end{eqnarray}

\begin{figure*}[!t]
      \begin{center}
\includegraphics[scale=0.9]{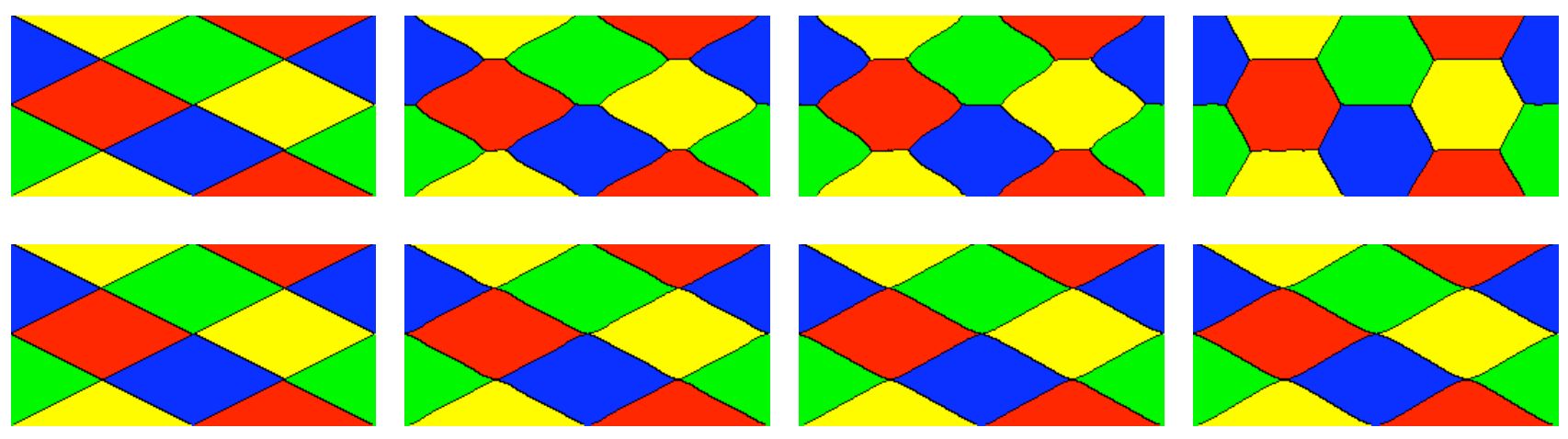}
\end{center}
\caption{Evolution of the field in the pentavac model, at times $t = 0, 10, 20, 30$ (left to right) for $\epsilon = 0.1, 0.03$ (top to bottom) and initial internal angle $\alpha = 53^{\circ}$. Each colour represents one of the vacua in the theory. For $\epsilon = 0.1$ the \xtype junction with $\alpha = 53^{\circ}$ is not the lowest energy state and it breaks apart into two \ytype junctions. In the $\epsilon = 0.03$ case the \xtype junction is the energetically preferred state and the \xtype junction retains its form.}\label{fig:pent-ep01_alpha53}
\end{figure*}

In \fref{fig:pent-ep01_alpha53} we present the evolution of the field for an initial configuration with an \xtype  junction having an initial internal angle $\alpha = 53^{\circ}$ (achieved by setting the number of horizontal and vertical grid-points to $N_x = 256, N_y = 128$, respectively). The case of $\epsilon = 0.1$ is in the top row. The \xtype junction clearly breaks apart into two \ytype junctions. However, the $\epsilon = 0.03$ case presented in the bottom row, the \xtype junction retains its form. Clearly $\qsubrm{\alpha}{c} = 53^{\circ}$ for some value of $\epsilon$ between $\epsilon = 0.1$ and $\epsilon = 0.03$.

\begin{figure*}[!t]
      \begin{center}
\includegraphics[scale=0.8]{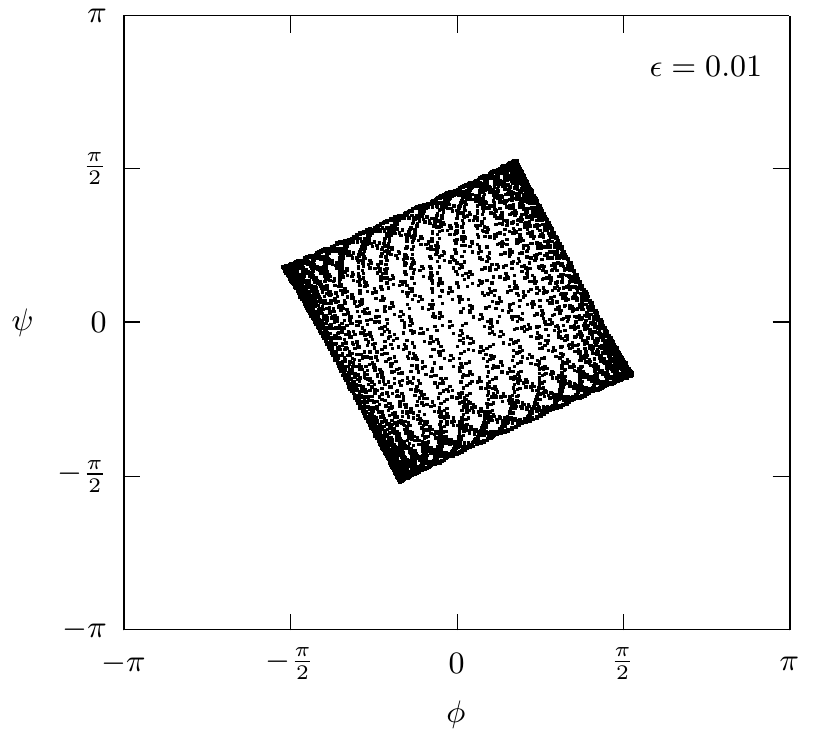}\qquad
\includegraphics[scale=0.8]{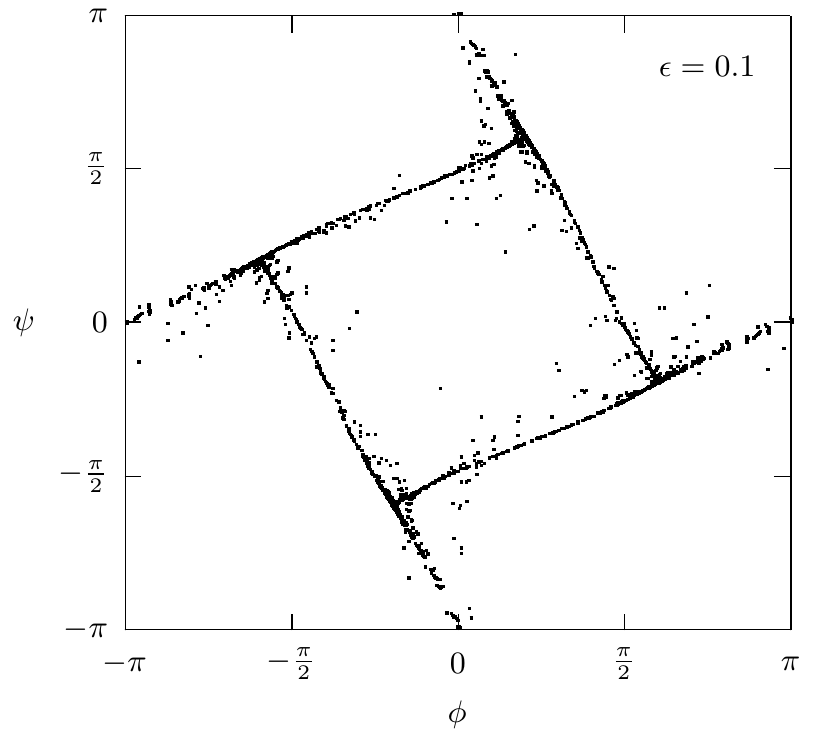}
\end{center}
\caption{The phases of the fields, $(\phi, \psi)$, evaluated at each point in space when $t=30$ following relaxation of a configuration having an initial \xtype junction with $\alpha = 53^{\circ}$ for  two different values of the symmetry breaking parameter $\epsilon$. The case of $\epsilon = 0.01$ has retained its form as an \xtype junction, while the case of $\epsilon = 0.1$ has relaxed into a \ytype junction configuration. One can clearly see that one crosses the origin in the $\epsilon = 0.01$ case, which correspond to high tension walls, whereas this does not take place in the $\epsilon = 0.1$ case with all walls being low tension. }\label{fig:pent_phase-eval_high_low}
\end{figure*}

A plot of the distribution of the phases $(\phi, \psi)$ at a given time will show where the points are located on the vacuum manifold, and will highlight any fundamental differences between systems whose \xtype junctions have and have not decayed. Referring back to the vacuum manifold in \fref{penta-vac-schematic-eppos}, we note that if any phases are found across the origin then high tension walls exist. Conversely, if this is not the case no high tension walls exist between opposing vacua and adjacent vacua must wrap the torus to produce high tension walls. We present the phases for $\epsilon = 0.1$ and $\epsilon = 0.01$ in \fref{fig:pent_phase-eval_high_low}, for an initial \xtype junction with $\alpha = 53^{\circ}$. One can clearly observe that there exists high tension walls in $\epsilon = 0.01$ case, and only low tension walls in the $\epsilon = 0.1$ case. These images represent the final state of the system. 

\begin{figure*}[!t]
      \begin{center}
\includegraphics[scale=0.9]{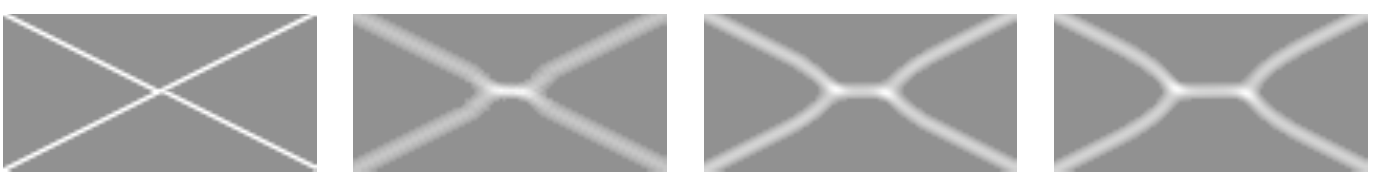}
\end{center}
\caption{Images of the potential energy density at times $t = 0, 5, 10, 15$ (left to right) for a decaying $\epsilon = 0.1$ \xtype junction. The value of the potential energy at a given location is denoted  by the relative brightness of the colouring. One observes a high tension wall initially being generated, but this swiftly decays into a low tension wall. }\label{fig:decay_pot_density_pentavac}
\end{figure*}

The potential energy density isosurfaces for the case of $\epsilon = 0.1$ are presented in \fref{fig:decay_pot_density_pentavac}. The value of the potential energy density is denoted by the relative brightness at a given location. We see that in the second image a high tension wall has formed, which subsequently decays to a low tension wall. This is compatible with the schematic decay process depicted in \fref{fig:xy}(b).  At first sight it appears that there is some sort of discontinuous jump as an \xtype decays into a \ytype junction: the field trajectories transition from wrapping around the toroidal vacuum to not wrapping. This transition is  energetically driven, and is therefore not a cause for concern.

We have developed an automated algorithm to compute the critical angle, $\qsubrm{\alpha}{c}$, at which an \xtype junction becomes unstable as a function of $\epsilon$.  We first compute the length and width of the wall for the junction at the centre of the grid, $L$ and $\delta$ respectively. If the junction is of \xtype  then the walls will be of comparable length, $L / \delta \approx 1$; however, if the junction is of \ytype then the horizontal wall will be much longer than the vertical wall, $L / \delta \gg 1$. We compute the value of $L/\delta$ after simulation time $t =\tau$, for initial conditions with an internal intersection angle $\alpha$. If $L/\delta < \zeta $, the junction is deemed to be of \xtype  and the vertical size of the box is reduced to lower $\alpha$. This process  continued until a \ytype junction with $L/\delta \geq \zeta $ is found. We take $\tau = 30$ for $N_x = 256$, and $\zeta = 2.5$ whose value was determined by convergence tests. 

\begin{figure*}[!t]
      \begin{center}
\includegraphics[scale=0.9]{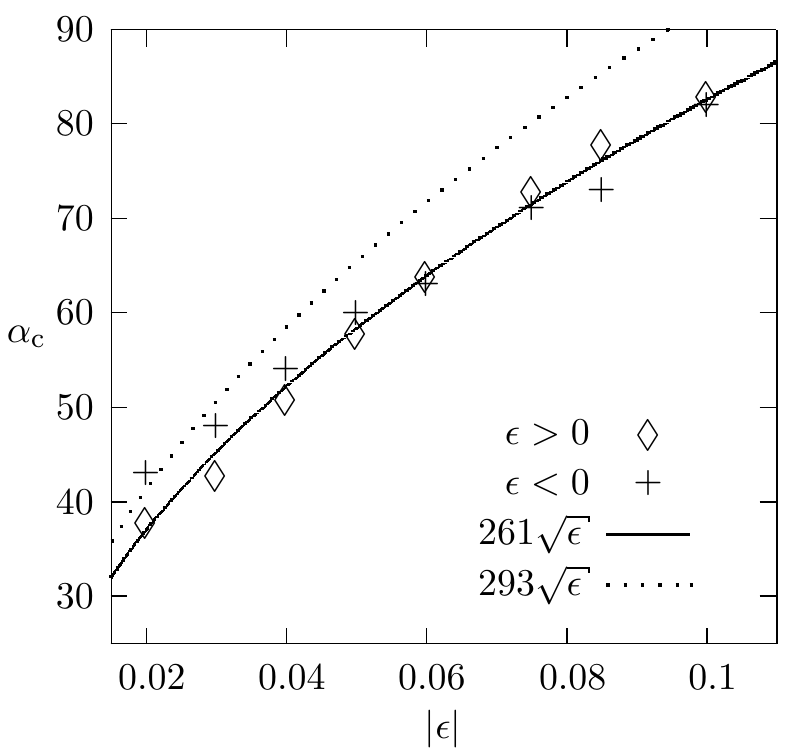}
\end{center}
\caption{The values of $\qsubrm{\alpha}{c}$ (in degrees) deduced numerically for various symmetry breaking parameters $\epsilon$ in the pentavac model. The solid line is the best fit to the data, whereas the dotted line is that predicted by the theory presented in the first section. }\label{plot:pent_x-junc-alg}
\end{figure*}

We present numerical estimates of $\qsubrm{\alpha}{c}$ in  \fref{plot:pent_x-junc-alg} for a range of positive and negative values of the symmetry breaking parameter. Our algorithm suffers from resolution effects as $\epsilon \rightarrow 0$ since the domain wall width widens, and therefore we only show results for $|\epsilon| \geq 0.02$.  The results show that: (1) $\qsubrm{\alpha}{c}$ is largely independent of the sign of $\epsilon$, something which is understood by the structure of the vacuum manifold (the manifolds for $\epsilon >0$ and $\epsilon <0$ have identical structure); (2) the numerical results have the same dependance on $\epsilon$ as the analytical prediction, however, their values are  $\sim10\%$ lower which can be understood as a rather trivial numerical resolution effect -- the calculation predicts the value of $\qsubrm{\alpha}{c}$ for the moment an \xtype junction breaks apart by an infinitesimal distance whereas the numerical calculation has to wait until $L/\delta = 2.5$. Unfortunately we are unable to reduce $\zeta$ any more due to numerical resolution effects.
 
We note that it is not possible within our framework to perform the reverse process. It is not energetically possible to  start off with two \ytype junctions and for them to contract into a high tension wall.

\section{Scaling dynamics from random initial conditions}
\label{sec:pentavac-scaling}
In the previous section we have shown that both \xtype and \ytype junctions can be stable, dependant on the intersection angle $\alpha$ and $\epsilon$. We now develop this further by testing how the dynamics of random domain wall networks are affected by varying the symmetry breaking parameter $\epsilon$. From our investigation of idealized \xtype junctions in the previous section we have seen that a lower $\epsilon$ implies that an \xtype junction is stable for a wider range of intersection angles, for example if $\epsilon = 0.01$ then \xtype junctions will be stable for $\alpha > 30^{\circ}$. Performing simulations of a random network will allow us to put a measure on the probability distribution of $\alpha$ in realistic situations.

\begin{figure*}[t]
      \begin{center}
	\includegraphics[scale=1.0]{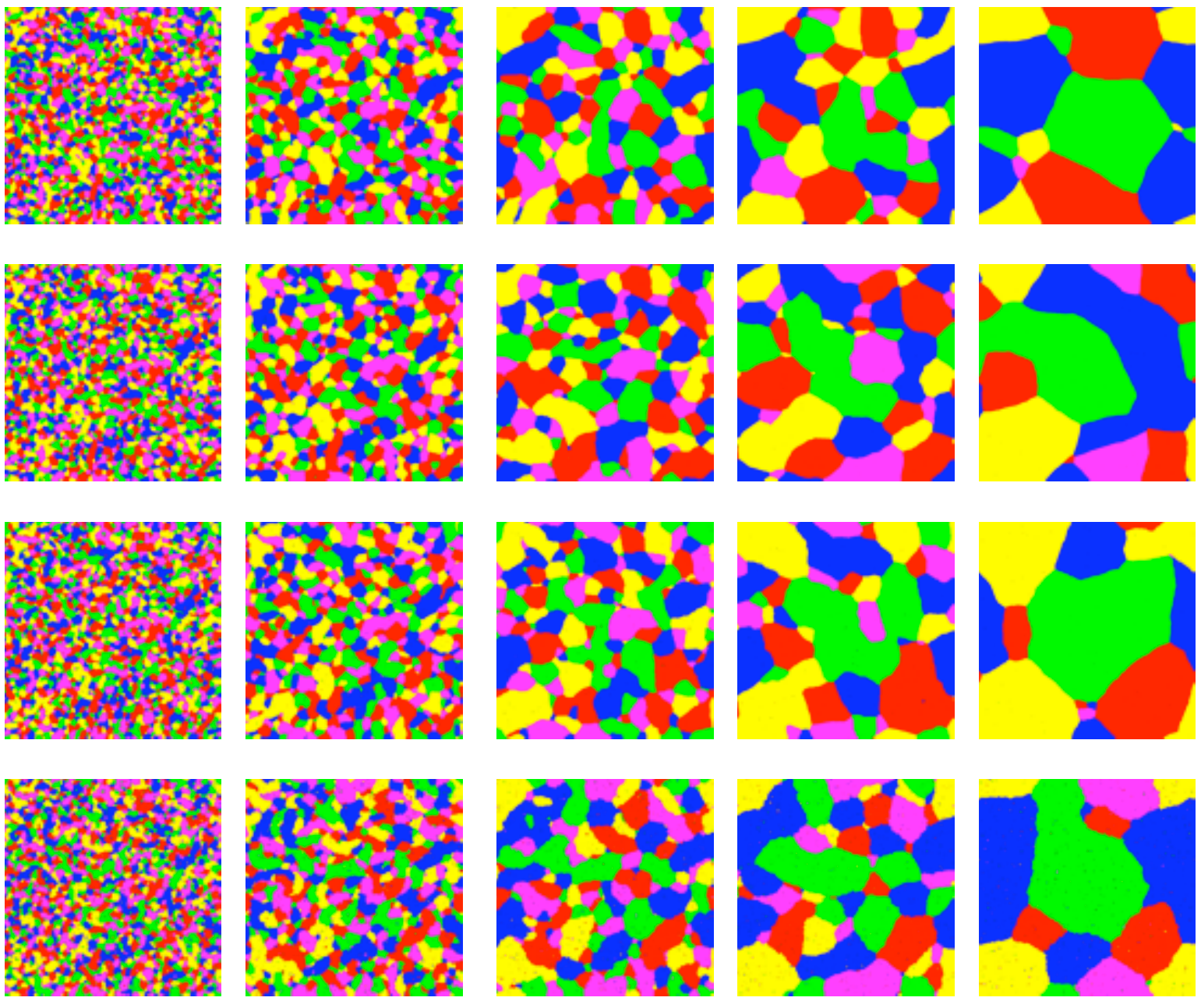}
	      \end{center}
\caption{Evolution of fields in the pentavac model, for $\epsilon = 0.1, 0.05, 0.03, 0.01$ (top to bottom) from random initial conditions on a $P=4096$ square grid. Images are at $t = 80, 160, 320, 640, 1280$ (left to right) and the colours correspond to the vacua of the pentavac theory.  Each colour represents one of the vacua in the theory, and are assigned by finding which of the vacua the field is closest to at each location.}\label{fig:pent_4096_scaling}
\end{figure*}

\begin{figure*}[t]
      \begin{center}
	\includegraphics[scale=1.0]{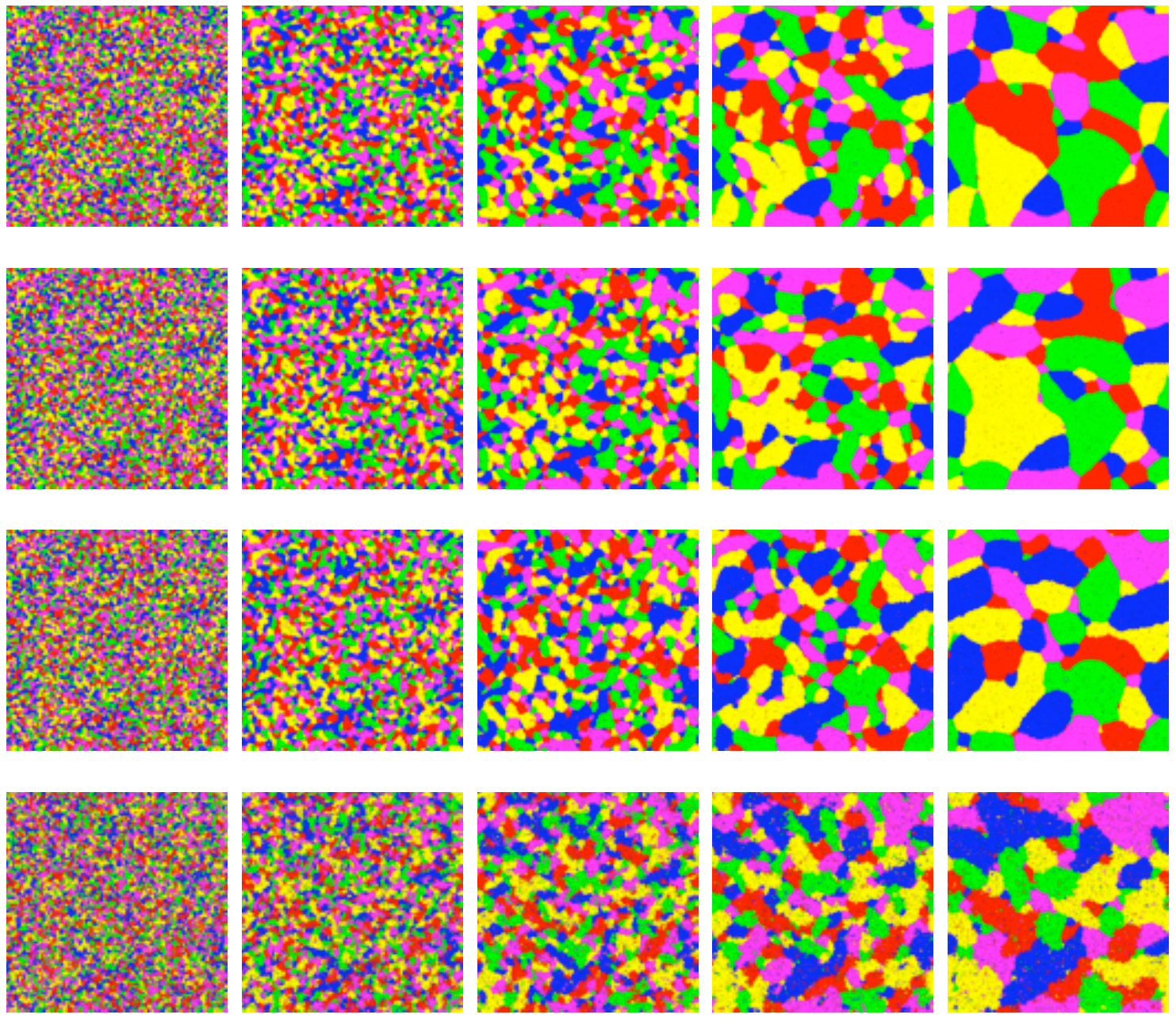}
      \end{center}
\caption{Evolution of the fields in the pentavac model, for systems with $\epsilon = 0.01, 0.008, 0.005, 0.001$ (top to bottom) at times $t = 80, 160, 320, 640, 1280$ (left to right), on a grid with  $P = 4096$ lattice sites and space step-size $\Delta x = 1.0$. }\label{fig:pent_smallep_images}
\end{figure*}

We have evolved the  equations of motion in Minkowski space with random initial domain occupation on lattices with $P = 4096$ grid-points in each direction, and space step-size $\Delta x = 0.5$ and an initial correlation length of one grid-square. We apply a damping term of magnitude 0.5 for the first 200 time-steps to smooth out unphysical discontinuities that result from our initial conditions.  We present  images of the resulting field configurations as a function of time in \fref{fig:pent_4096_scaling} and \fref{fig:pent_smallep_images} for various values of $\epsilon$. The ``mottled'' appearance of the $\epsilon \leq 0.01$ simulations is due to the rather shallow potential wells defining the domains as $\epsilon \rightarrow 0$; however domains still clearly form. By carefully inspecting the field images we have been able to find that there is a higher fraction of \xtype junctions to \ytype junctions in simulations with lower $\epsilon$; however, the \xtype junctions that are present at early time do not seem to survive to late time.

\begin{figure*}[!h]
      \begin{center}   
\subfigure[\, Evolution of the number of domain walls, $\qsubrm{N}{dw}(t)$.]{\includegraphics[scale=0.28]{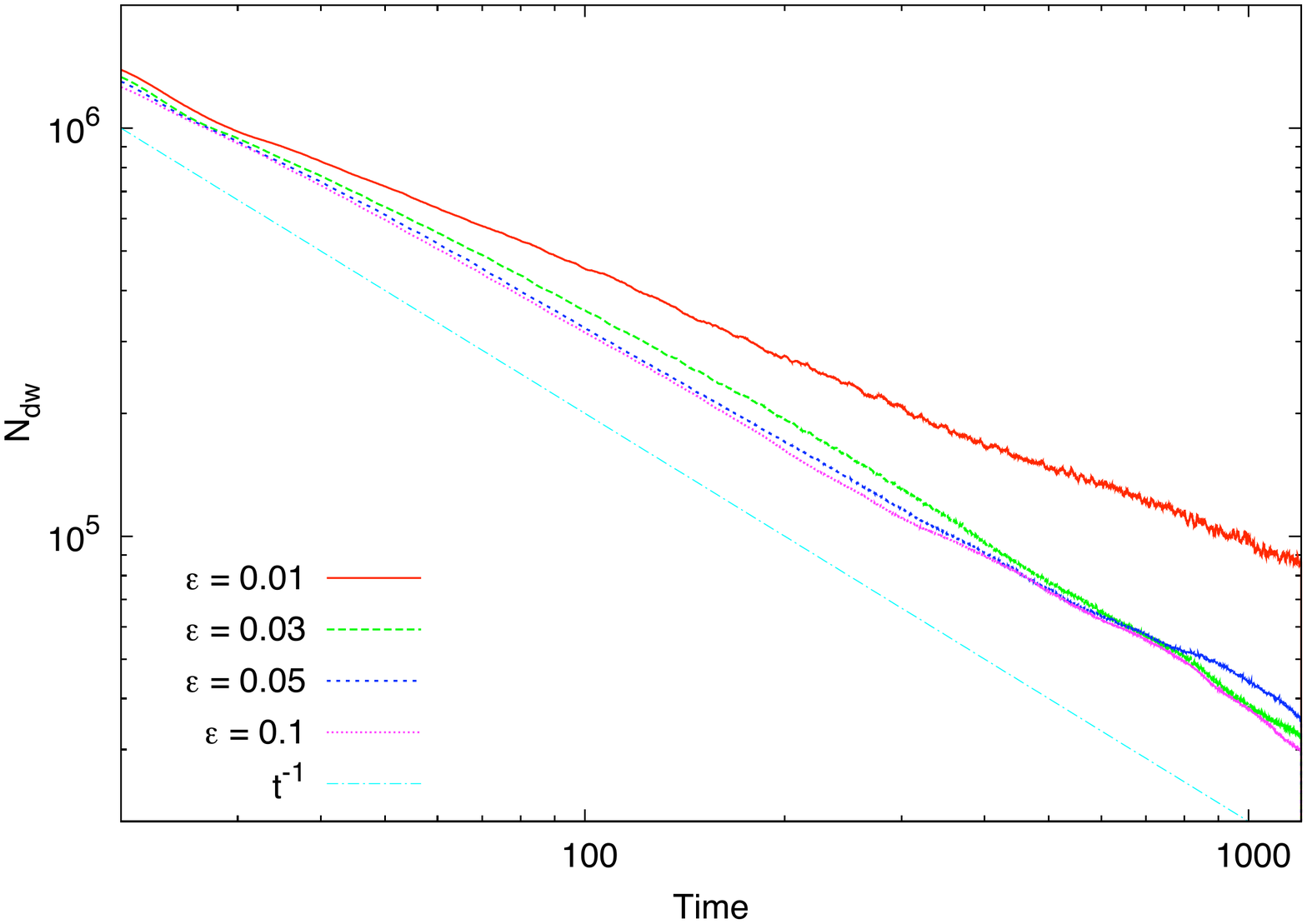}\qquad\includegraphics[scale=0.28]{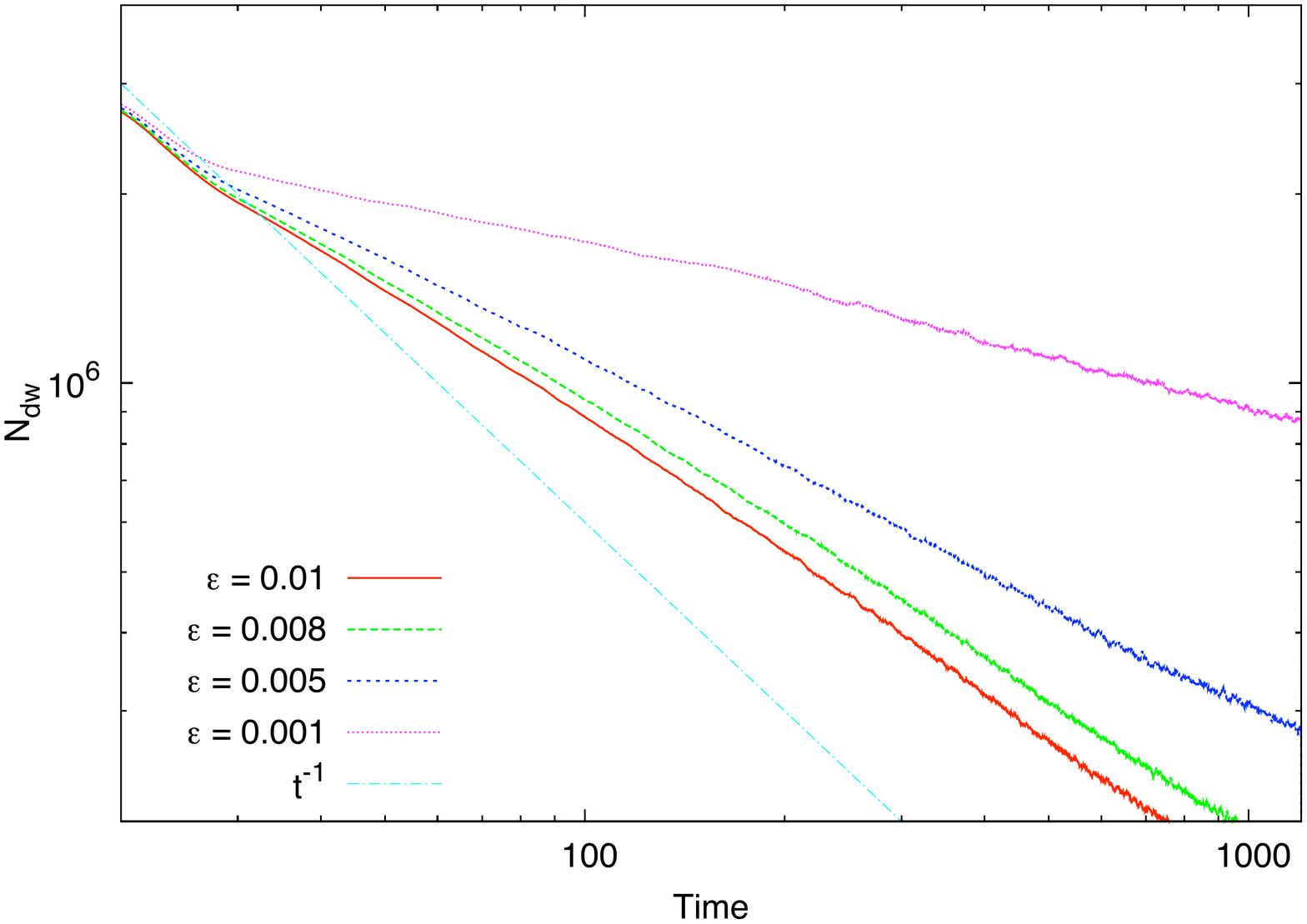}}
\subfigure[\, Evolution of the scaling exponents, $\gamma(t)$.]{\includegraphics[scale=0.28]{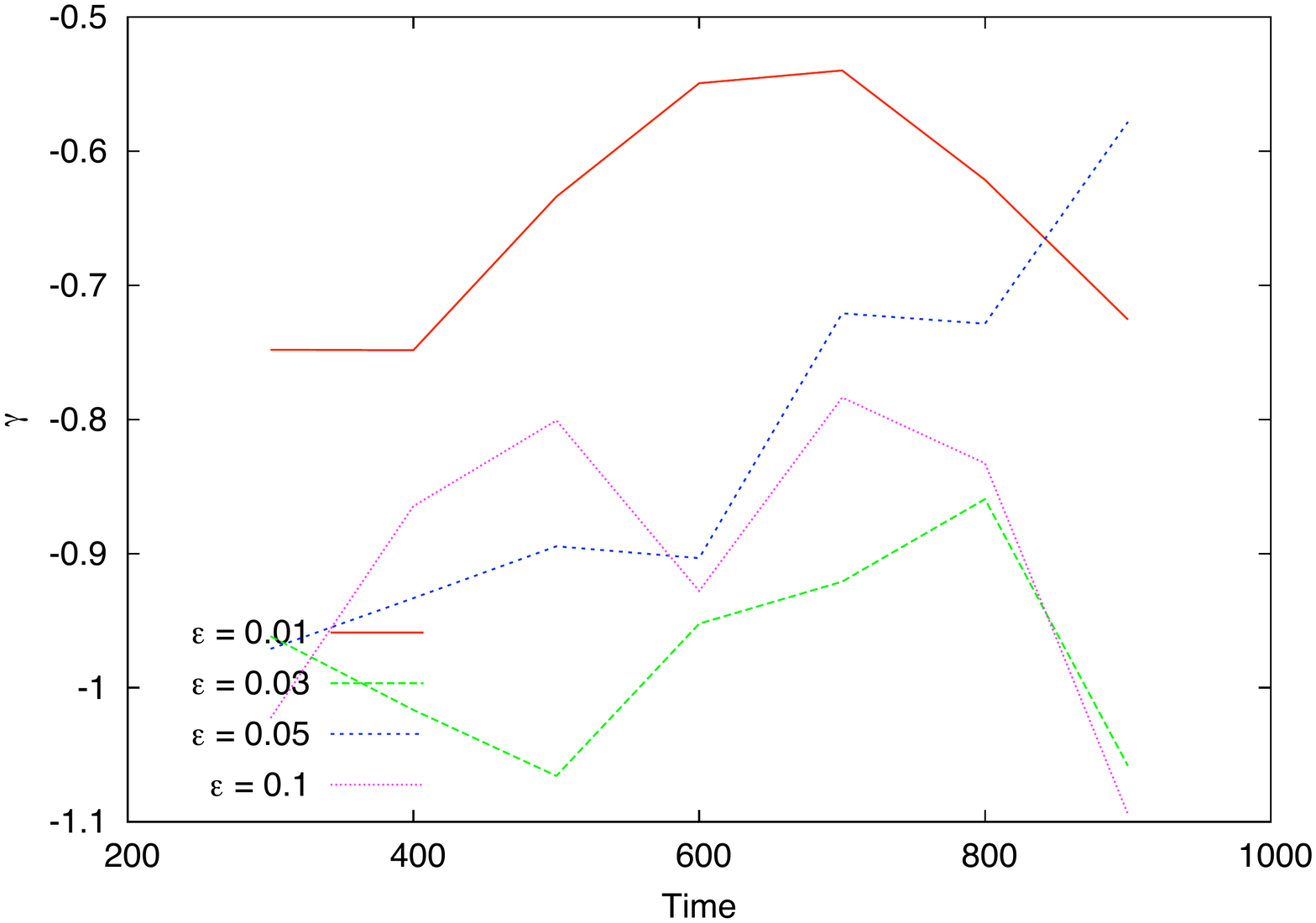}\qquad\includegraphics[scale=0.28]{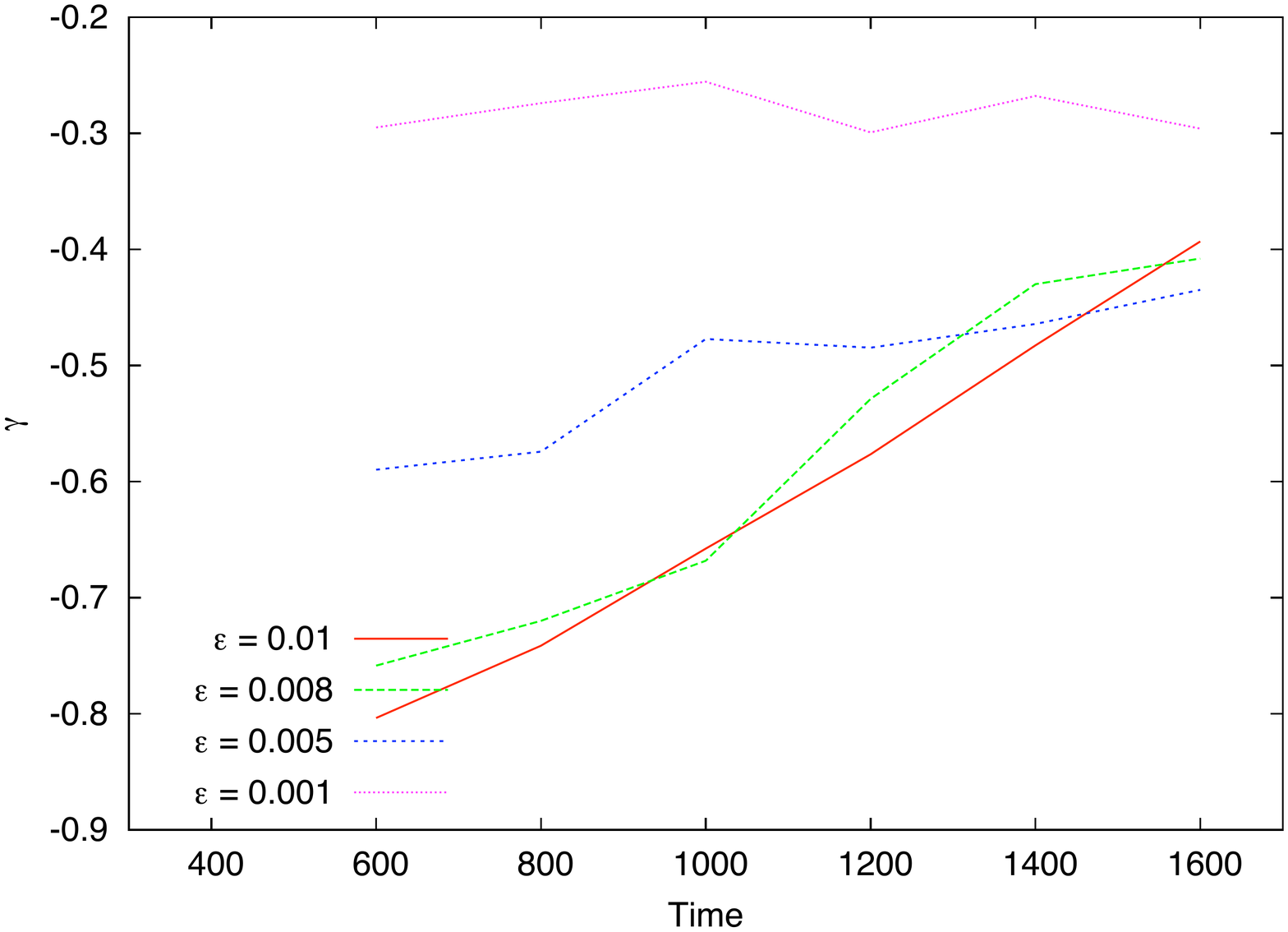}}
      \end{center}
\caption{Evolution of the number of domain walls $\qsubrm{N}{dw}$ and scaling exponents $\gamma$ from random initial conditions, in the pentavac model, for various symmetry breaking parameters $\epsilon$, where the grid size is $P = 4096$.  In the left-hand panels, $\Delta x = 0.5$ is used, with scaling exponents being computed in bins of 100 units of time. In the right-hand panels, $\Delta x = 1.0$ is used and scaling exponents are computed in time-bins of 200. One can easily observe that the scaling dynamic $\qsubrm{N}{dw}(t)$ is different in the cases $\epsilon >0.01$ and $\epsilon \leq 0.01$. The exponent, $\gamma(t)$, of the power-law that $\qsubrm{N}{dw}(t)$ satisfies  clearly changes as $\epsilon$ changes, with $\gamma \rightarrow 0$ as $\epsilon \rightarrow 0$.  }\label{fig:pent_4096_scaling-nwalls}     
\end{figure*}

To understand how the wall network evolves we calculate the number of domain walls at every time-step and build up a plot of the evolution of the number of domain walls $\qsubrm{N}{dw}(t)$ and compute the scaling exponents $\gamma$ defined by $\qsubrm{N}{dw} \propto t^{\gamma}$. A domain wall is numerically defined when the vacuum that the field occupies at a given lattice site is different from the vacuum occupied by any of its neighboring lattice sites. The standard lore is $\gamma = -1$ \cite{PhysRevD.68.103506, Avelino20051, Avelino200763, PhysRevD.74.023528, PhysRevD.78.103508} in which case the network collapses as fast as causality allows. Deviation from this value signals the existence of some force resisting the collapse of the wall network and the possible formation of a lattice.

We plot the evolution of the number of domain walls for the systems in \fref{fig:pent_4096_scaling} and \fref{fig:pent_smallep_images}  in \fref{fig:pent_4096_scaling-nwalls}(a). For $\epsilon \geq 0.03$ the number of domain walls follows the standard $t^{-1}$ power law. However there is substantial deviation for $\epsilon \leq 0.01$. Plotting the evolution of the scaling exponents $\gamma(t)$ in \fref{fig:pent_4096_scaling-nwalls}(b) confirms this.

\begin{figure*}[!t]
      \begin{center}
\subfigure[\, Fixed physical size]{\includegraphics[scale=0.28]{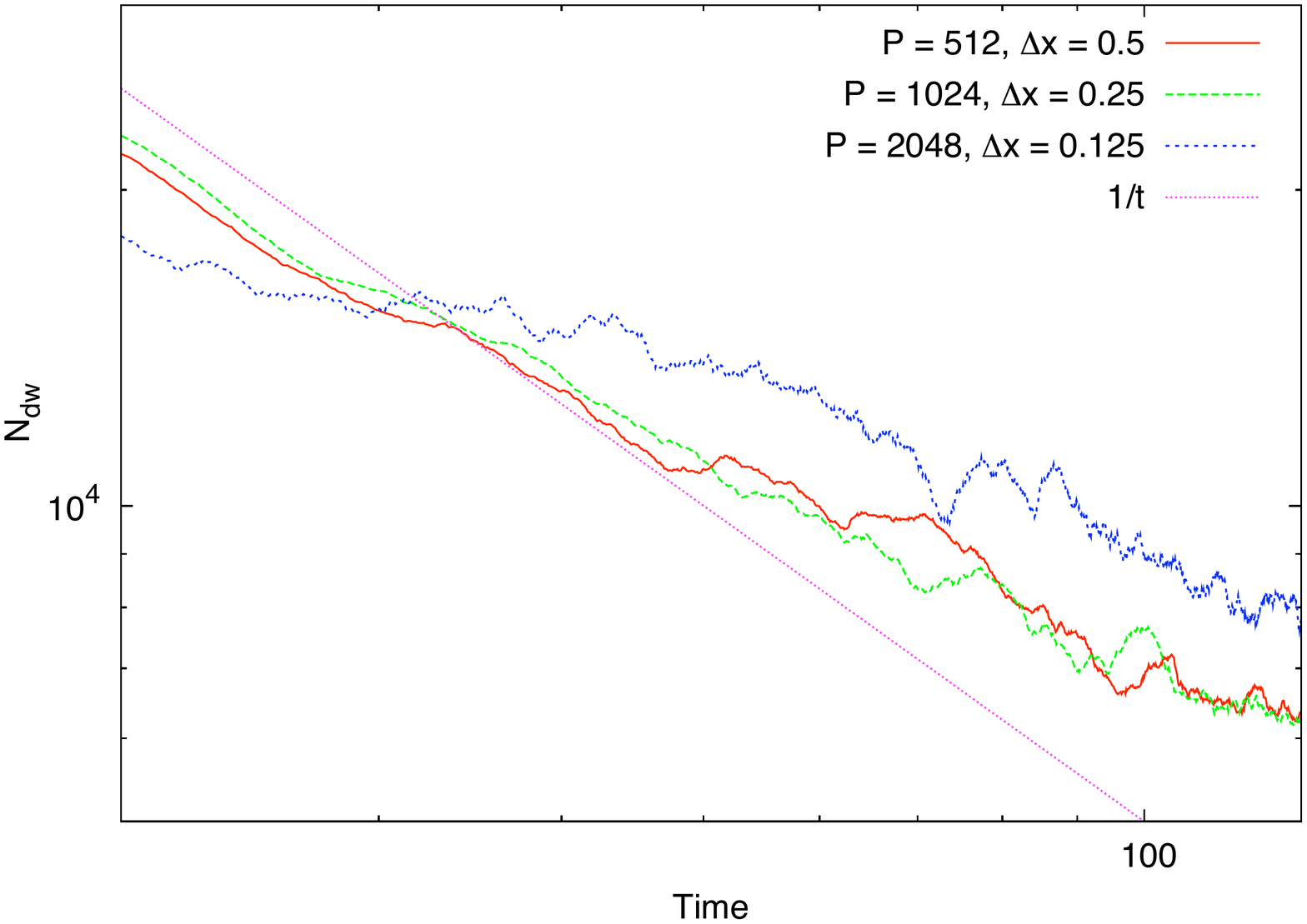}}
\subfigure[\, Fixed $P = 2048$]{\includegraphics[scale=0.28]{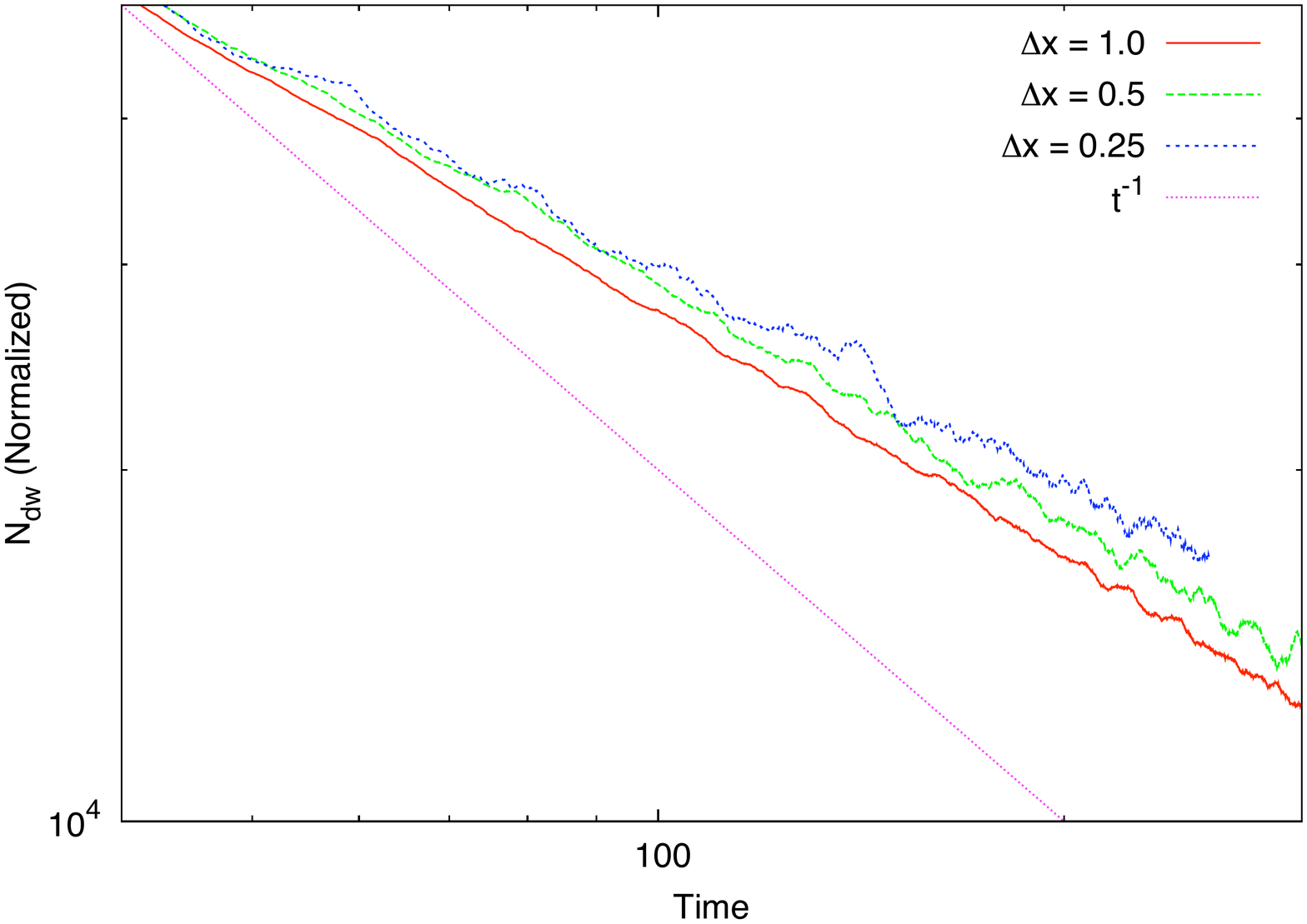}}
      \end{center}
\caption{Evolution of the number of domain walls with $\epsilon = 0.01$, with varying resolution: (a)   $P = 512, 1024, 2048$ with varying space step-size $\Delta x = 0.5, 0.25, 0.125$ and initial correlation lengths $\xi = 1,2,4$ respectively; (b)  $P = 2048$ and varied $\Delta x$, keeping $\xi = 1$.}\label{fig:nwalls_comp_resolution}
\end{figure*}

One might be concerned that this effect is due to the finite resolution and dynamic range of the simulations particularly since the width of the domain wall is expected to be $\qsubrm{\delta}{w} \propto \epsilon^{-\half}$. To test our results we have performed  a number of convergence tests. Simulations with various space step sizes, correlation lengths and a variety of different physical sizes of grids were used. A correlation length $\xi$ is imposed by making the system have blocks of $\xi \times \xi$ grid-points initially in the same vacuum.
We will focus on the case of $\epsilon = 0.01$ since it is close to the parameter range where the non-standard behavior starts to become noticeable. First we evolved with $P = 512, 1024, 2048$ with correlation lengths $\xi = 1, 2, 4$ and space step-sizes $\Delta x = 0.5, 0.25, 0.125$ respectively; these fix the physical size of the box while varying its resolution. Secondly we change the physical size but retain the number of grid-points. To ensure that the Courant condition is satisfied we changed the time step-size for systems with $\Delta x = 0.125$ to $\Delta t = 0.05$. In \fref{fig:nwalls_comp_resolution} we show a plot of the evolution of the number of domain walls for these test systems. All are consistent with the same deviation from $\gamma = -1$ shown in \fref{fig:pent_4096_scaling}, and all have qualitatively the same scaling dynamics.

A separate concern is that the effect is due to the energy that was originally localized on the wall network as walls collapse. As previously pointed out the energy required to created a wall is very low for small $\epsilon$. As a wall network collapses energy becomes released from the walls into radiation which may be sufficient to create new ``secondary'' walls. As the original wall network collapses   the amount of radiation increases with a consequent  increase in the rate of ``secondary'' wall production. Such a mechanism could account for the deviation from the standard scaling behavior we observed. 

\begin{figure*}[!t]
      \begin{center}
\subfigure[\, $\epsilon = 0.1$]{\includegraphics[scale=0.28]{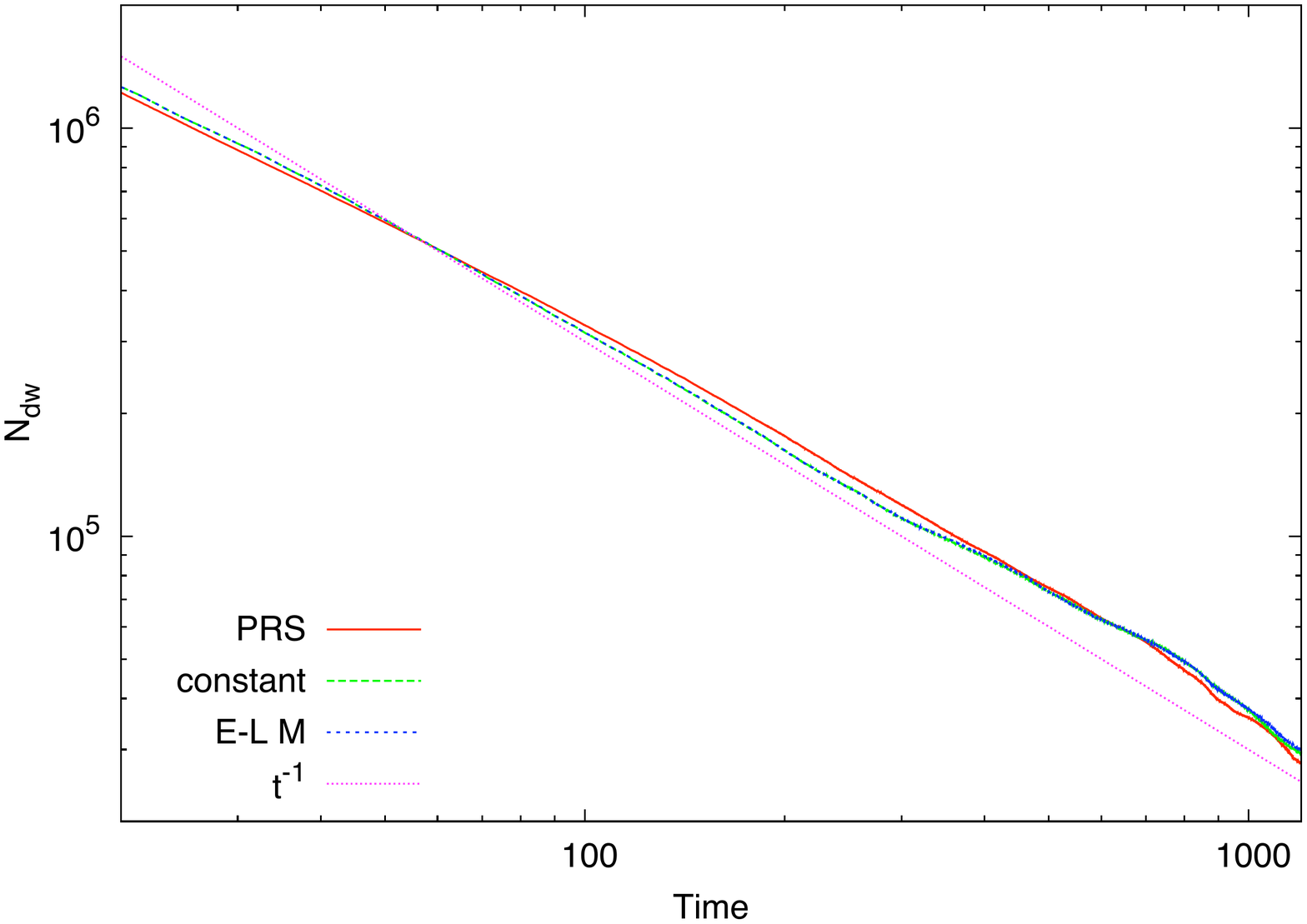}}
\subfigure[\, $\epsilon = 0.01$]{\includegraphics[scale=0.28]{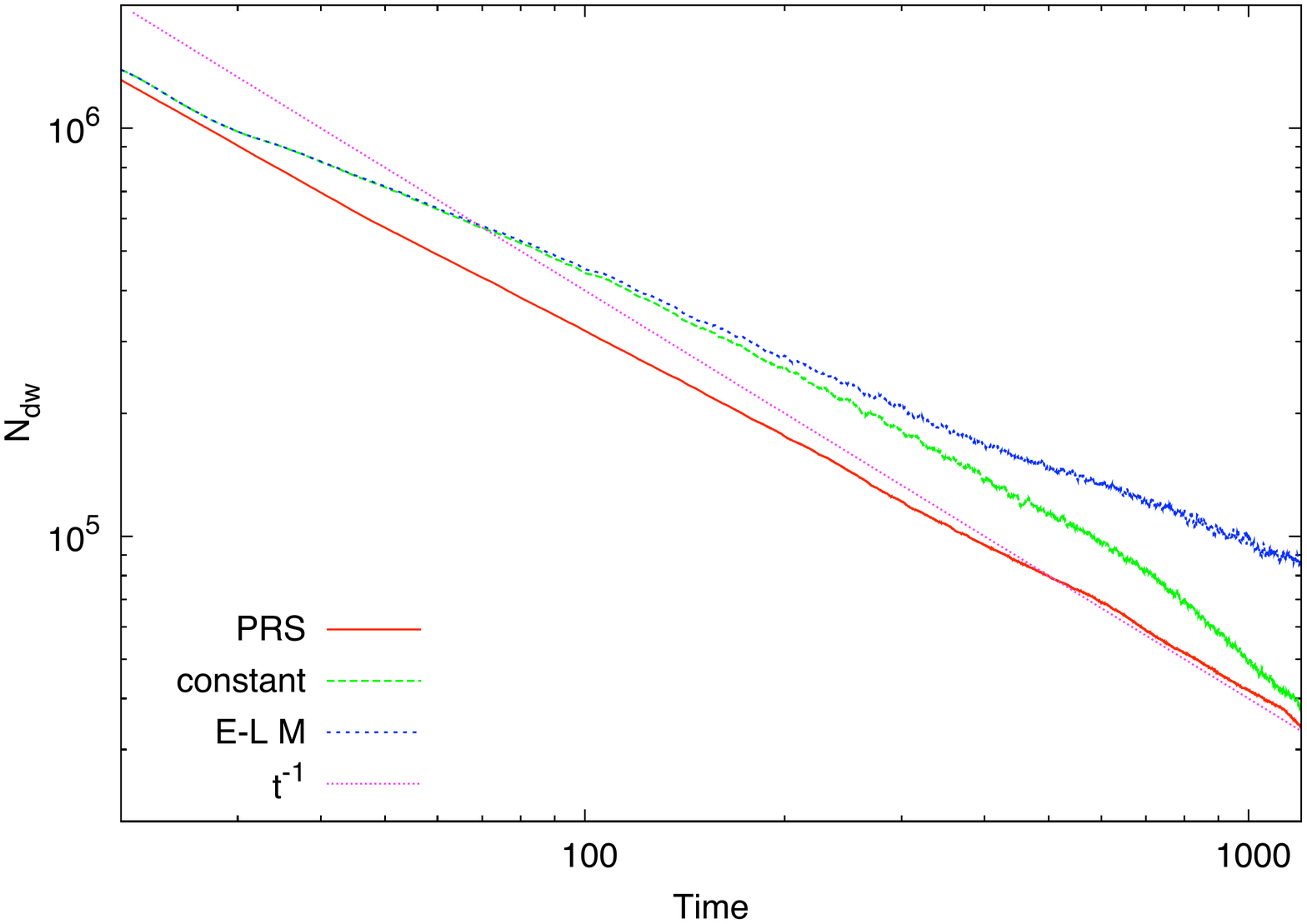}}
      \end{center}
\caption{Evolution of the number of domain walls for three ``different'' evolution algorithms with $\epsilon = 0.1, 0.01$; we use $\Delta x = 0.5, \Delta t = 0.1, P = 4096$ and random initial conditions. The line denoted ``E-L M'' is the evolution of the standard Euler-Lagrange equations in Minkowski spacetime. The line denoted ``constant'' has a constant damping.  The line denoted ``PRS'' uses the PRS algorithm in the radiation era. The implementation of these algorithms is given in (\ref{eq:sec:4.2-damping}).}\label{fig:nwall_prs_n0_elm}
\end{figure*}

If this ``background radiation'' can be removed by some means then the scaling dynamics may revert to the standard $t^{-1}$ law. This has been tested by evolving a modified set of field equations, where radiation is   removed. This can be done by including a constant damping term   (but very small) or by using a modified version of the algorithm by Press, Ryden and Spergel (PRS), where the damping term is time-dependant, which has been suggested as a way of modeling an expanding Universe \cite{1989ApJ...347..590P, avelino_dw_carter}. These equations of motion are of the form
\bea
\ddot{\phi} + \mathcal{D}\dot{\phi}- \nabla^2\phi + \frac{\dd V}{\dd\phi}=0.
\eea
In each of the modified evolution algorithms the coefficient $\mathcal{D}$ of the damping term takes on the following form:
\bea
\label{eq:sec:4.2-damping}
\qsubrm{\mathcal{D}}{PRS}= \frac{3}{t}, \qquad \qsubrm{\mathcal{D}}{constant} = \left\{ \begin{array}{ccc} \mathcal{D}_0 && t < 20,\\ 10^{-3}\mathcal{D}_0 && t >20.
\end{array}\right.
\eea
We take $\mathcal{D}_0 = 0.5$.
The Euler-Lagrange equations in Minkowski spacetime have $\mathcal{D}  = 0.5$ for $t < 20$ and $\mathcal{D} =0$ thereafter.
 In \fref{fig:nwall_prs_n0_elm} we compare the evolution of the number of domain walls in these ``constant damping'' and PRS algorithms (as well as with the standard Euler-Lagrange equations in Minkowski spacetime). It is clear that in the $\epsilon = 0.01$ simulation the evolution is substantially different in these damped algorithms to the evolution in the Euler-Lagrange equations in Minkowski spacetime.

Because the inclusion of a dissipative term restores the $t^{-1}$ scaling law (at least for a substantial period of time) it is no longer obvious which evolution algorithm corresponds to  results relevant in cosmology. This issue may be resolved by an extensive systematic study of the results of the equations of motion for a range of models with as large a dynamic range as feasible.

\section{Conclusions}
\label{ref:conc}
In this paper we have furthered the understanding of the stability of \xtype and \ytype junctions. We have calculated that if an \xtype junction has an internal intersection angle $\alpha$, then if $\alpha > \qsubrm{\alpha}{c}$ the \xtype junction is stable, whereas if $\alpha < \qsubrm{\alpha}{c}$ the \xtype junction decays into two \ytype junctions. In Carter's pentavac model the dependance on the critical intersection angle $\qsubrm{\alpha}{c}$ upon the models symmetry breaking parameter $\epsilon$ has been derived to be
\bea
\qsubrm{\alpha}{c} = 293^{\circ}\sqrt{|\epsilon|}.
\eea
The validity of this calculation has been verified by numerical experimentation, and is found to hold regardless of the sign of $\epsilon$. 

A consequence of this relationship is that if $\epsilon$ is relatively large (for example, $\epsilon = 0.05$) the \xtype junctions cannot withstand much ``squashing'' before they break apart; they can only withstand squashing to an angle of $\alpha \approx 65^{\circ}$. Therefore,   \xtype junctions  cannot be expected to survive in physically relevant systems. This is what was observed by Avelino et al \cite{avelino_dw_carter} but they expanded the statement to include all systems with \xtype junctions. However, if $\epsilon$ is taken to be small (for example $\epsilon = 0.01$) \xtype junctions are stable for intersection angles  above $\alpha \approx 30^{\circ}$ and therefore could be expected to survive in physically relevant systems. The ability of an \xtype junction to retain its form under much more extreme intersections is increased as $\epsilon$ is decreased. 

Whilst \xtype junctions become more frequent in simulations from an initially random configuration of vacua when $\epsilon$ is small, the \xtype junctions do not survive to late time. For systems with $\epsilon \leq 0.01$   the scaling dynamics of the resulting network becomes modified compared to that of a system with   $\epsilon >0.01$. The amount by which the evolution is modified was clearly shown to depend on the value of $\epsilon$ and we believe that this effect is due to ``background radiation'' being sufficient to create walls.

\section*{Acknowledgements}
We have benefited from code written by Chris Welshman which formed the basis of our visualization software. RB thanks Brandon Carter for impetus and many useful conversations.



\begin{thebibliography}{25}
\expandafter\ifx\csname natexlab\endcsname\relax\def\natexlab#1{#1}\fi
\expandafter\ifx\csname bibnamefont\endcsname\relax
  \def\bibnamefont#1{#1}\fi
\expandafter\ifx\csname bibfnamefont\endcsname\relax
  \def\bibfnamefont#1{#1}\fi
\expandafter\ifx\csname citenamefont\endcsname\relax
  \def\citenamefont#1{#1}\fi
\expandafter\ifx\csname url\endcsname\relax
  \def\url#1{\texttt{#1}}\fi
\expandafter\ifx\csname urlprefix\endcsname\relax\def\urlprefix{URL }\fi
\providecommand{\bibinfo}[2]{#2}
\providecommand{\eprint}[2][]{\url{#2}}

\bibitem[{\citenamefont{Copeland et~al.}(2004)\citenamefont{Copeland, Myers,
  and Polchinski}}]{1126-6708-2004-06-013}
\bibinfo{author}{\bibfnamefont{E.}~\bibnamefont{Copeland}},
  \bibinfo{author}{\bibfnamefont{R.}~\bibnamefont{Myers}}, \bibnamefont{and}
  \bibinfo{author}{\bibfnamefont{J.}~\bibnamefont{Polchinski}},
  \bibinfo{journal}{Journal of High Energy Physics}
  \textbf{\bibinfo{volume}{2004}}, \bibinfo{pages}{013} (\bibinfo{year}{2004}).

\bibitem[{\citenamefont{Saffin}(2005)}]{1126-6708-2005-09-011}
\bibinfo{author}{\bibfnamefont{P.}~\bibnamefont{Saffin}},
  \bibinfo{journal}{Journal of High Energy Physics}
  \textbf{\bibinfo{volume}{2005}}, \bibinfo{pages}{011} (\bibinfo{year}{2005}).

\bibitem[{\citenamefont{Bevis and Saffin}(2008)}]{PhysRevD.78.023503}
\bibinfo{author}{\bibfnamefont{N.}~\bibnamefont{Bevis}} \bibnamefont{and}
  \bibinfo{author}{\bibfnamefont{P.}~\bibnamefont{Saffin}},
  \bibinfo{journal}{Phys. Rev. D} \textbf{\bibinfo{volume}{78}},
  \bibinfo{pages}{023503} (\bibinfo{year}{2008}).

\bibitem[{\citenamefont{Bevis et~al.}(2009)\citenamefont{Bevis, Copeland,
  Martin, Niz, Pourtsidou, Saffin, and Steer}}]{PhysRevD.80.125030}
\bibinfo{author}{\bibfnamefont{N.}~\bibnamefont{Bevis}},
  \bibinfo{author}{\bibfnamefont{E.}~\bibnamefont{Copeland}},
  \bibinfo{author}{\bibfnamefont{P.}~\bibnamefont{Martin}},
  \bibinfo{author}{\bibfnamefont{G.}~\bibnamefont{Niz}},
  \bibinfo{author}{\bibfnamefont{A.}~\bibnamefont{Pourtsidou}},
  \bibinfo{author}{\bibfnamefont{P.}~\bibnamefont{Saffin}}, \bibnamefont{and}
  \bibinfo{author}{\bibfnamefont{D.~A.} \bibnamefont{Steer}},
  \bibinfo{journal}{Phys. Rev. D} \textbf{\bibinfo{volume}{80}},
  \bibinfo{pages}{125030} (\bibinfo{year}{2009}).

\bibitem[{\citenamefont{{D. Thouless}}(1997)}]{Thouless}
\bibinfo{author}{\bibnamefont{{D. Thouless}}},
  \emph{\bibinfo{title}{{\textit{Topological quantum numbers in nonrelativistic
  physics}}}} (\bibinfo{publisher}{{World Scientific}}, \bibinfo{year}{1997}).

\bibitem[{\citenamefont{{M. Bucher and D.N.
  Spergel}}(1999)}]{bucher_spergel_1999}
\bibinfo{author}{\bibnamefont{{M. Bucher and D.N. Spergel}}},
  \bibinfo{journal}{Phys.Rev} \textbf{\bibinfo{volume}{D60}},
  \bibinfo{pages}{043505} (\bibinfo{year}{1999}).

\bibitem[{\citenamefont{{R.A. Battye, M. Bucher and D.
  Spergel}}(1999)}]{battye_bucher_spergel_1999}
\bibinfo{author}{\bibnamefont{{R.A. Battye, M. Bucher and D. Spergel}}}
  (\bibinfo{year}{1999}), \bibinfo{note}{{arXiv: astro-ph/9908047}}.

\bibitem[{\citenamefont{Battye and Moss}(2007)}]{PhysRevD.76.023005}
\bibinfo{author}{\bibfnamefont{R.}~\bibnamefont{Battye}} \bibnamefont{and}
  \bibinfo{author}{\bibfnamefont{A.}~\bibnamefont{Moss}},
  \bibinfo{journal}{Phys. Rev. D} \textbf{\bibinfo{volume}{76}},
  \bibinfo{pages}{023005} (\bibinfo{year}{2007}).

\bibitem[{\citenamefont{Battye et~al.}(2006)\citenamefont{Battye, Chachoua, and
  Moss}}]{PhysRevD.73.123528}
\bibinfo{author}{\bibfnamefont{R.}~\bibnamefont{Battye}},
  \bibinfo{author}{\bibfnamefont{E.}~\bibnamefont{Chachoua}}, \bibnamefont{and}
  \bibinfo{author}{\bibfnamefont{A.}~\bibnamefont{Moss}},
  \bibinfo{journal}{Phys. Rev. D} \textbf{\bibinfo{volume}{73}},
  \bibinfo{pages}{123528} (\bibinfo{year}{2006}).

\bibitem[{\citenamefont{Avelino
  et~al.}(2006{\natexlab{a}})\citenamefont{Avelino, Martins, Menezes, Menezes,
  and Oliveira}}]{PhysRevD.73.123519}
\bibinfo{author}{\bibfnamefont{P.}~\bibnamefont{Avelino}},
  \bibinfo{author}{\bibfnamefont{C.~P.} \bibnamefont{Martins}},
  \bibinfo{author}{\bibfnamefont{J.}~\bibnamefont{Menezes}},
  \bibinfo{author}{\bibfnamefont{R.}~\bibnamefont{Menezes}}, \bibnamefont{and}
  \bibinfo{author}{\bibfnamefont{J.}~\bibnamefont{Oliveira}},
  \bibinfo{journal}{Phys. Rev. D} \textbf{\bibinfo{volume}{73}},
  \bibinfo{pages}{123519} (\bibinfo{year}{2006}{\natexlab{a}}).

\bibitem[{\citenamefont{Avelino
  et~al.}(2006{\natexlab{b}})\citenamefont{Avelino, Martins, Menezes, Menezes,
  and Oliveira}}]{PhysRevD.73.123520}
\bibinfo{author}{\bibfnamefont{P.}~\bibnamefont{Avelino}},
  \bibinfo{author}{\bibfnamefont{C.}~\bibnamefont{Martins}},
  \bibinfo{author}{\bibfnamefont{J.}~\bibnamefont{Menezes}},
  \bibinfo{author}{\bibfnamefont{R.}~\bibnamefont{Menezes}}, \bibnamefont{and}
  \bibinfo{author}{\bibfnamefont{J.}~\bibnamefont{Oliveira}},
  \bibinfo{journal}{Phys. Rev. D} \textbf{\bibinfo{volume}{73}},
  \bibinfo{pages}{123520} (\bibinfo{year}{2006}{\natexlab{b}}).

\bibitem[{\citenamefont{Battye et~al.}(2005)\citenamefont{Battye, Carter,
  Chachoua, and Moss}}]{PhysRevD.72.023503}
\bibinfo{author}{\bibfnamefont{R.}~\bibnamefont{Battye}},
  \bibinfo{author}{\bibfnamefont{B.}~\bibnamefont{Carter}},
  \bibinfo{author}{\bibfnamefont{E.}~\bibnamefont{Chachoua}}, \bibnamefont{and}
  \bibinfo{author}{\bibfnamefont{A.}~\bibnamefont{Moss}},
  \bibinfo{journal}{Phys. Rev. D} \textbf{\bibinfo{volume}{72}},
  \bibinfo{pages}{023503} (\bibinfo{year}{2005}).

\bibitem[{\citenamefont{Battye et~al.}(2009)\citenamefont{Battye, Pearson,
  Pike, and Sutcliffe}}]{BattyePearson-kvform}
\bibinfo{author}{\bibfnamefont{R.}~\bibnamefont{Battye}},
  \bibinfo{author}{\bibfnamefont{J.}~\bibnamefont{Pearson}},
  \bibinfo{author}{\bibfnamefont{S.}~\bibnamefont{Pike}}, \bibnamefont{and}
  \bibinfo{author}{\bibfnamefont{P.~M.} \bibnamefont{Sutcliffe}},
  \bibinfo{journal}{JCAP} \textbf{\bibinfo{volume}{0909}}, \bibinfo{pages}{039}
  (\bibinfo{year}{2009}), \eprint{0908.1865}.

\bibitem[{\citenamefont{Battye and Pearson}(2010)}]{BattyePearson-charge}
\bibinfo{author}{\bibfnamefont{R.}~\bibnamefont{Battye}} \bibnamefont{and}
  \bibinfo{author}{\bibfnamefont{J.}~\bibnamefont{Pearson}},
  \bibinfo{journal}{Phys. Rev. D} \textbf{\bibinfo{volume}{82}},
  \bibinfo{pages}{125001} (\bibinfo{year}{2010}).

\bibitem[{\citenamefont{Antunes et~al.}(2004)\citenamefont{Antunes, Pogosian,
  and Vachaspati}}]{PhysRevD.69.043513}
\bibinfo{author}{\bibfnamefont{N.~D.} \bibnamefont{Antunes}},
  \bibinfo{author}{\bibfnamefont{L.}~\bibnamefont{Pogosian}}, \bibnamefont{and}
  \bibinfo{author}{\bibfnamefont{T.}~\bibnamefont{Vachaspati}},
  \bibinfo{journal}{Phys. Rev. D} \textbf{\bibinfo{volume}{69}},
  \bibinfo{pages}{043513} (\bibinfo{year}{2004}).

\bibitem[{\citenamefont{{B. Carter}}(2005)}]{carter_pent_1}
\bibinfo{author}{\bibnamefont{{B. Carter}}}, \bibinfo{journal}{Int. J. Theor.
  Phys} \textbf{\bibinfo{volume}{44}}, \bibinfo{pages}{1729}
  (\bibinfo{year}{2005}).

\bibitem[{\citenamefont{{B. Carter}}(2008)}]{carter_x-stability}
\bibinfo{author}{\bibnamefont{{B. Carter}}}, \bibinfo{journal}{Class. Quant.
  Grav} \textbf{\bibinfo{volume}{25}}, \bibinfo{pages}{154001}
  (\bibinfo{year}{2008}).

\bibitem[{\citenamefont{Avelino et~al.}(2009)\citenamefont{Avelino, Oliveira,
  Menezes, and Menezes}}]{avelino_dw_carter}
\bibinfo{author}{\bibfnamefont{P.}~\bibnamefont{Avelino}},
  \bibinfo{author}{\bibfnamefont{J.}~\bibnamefont{Oliveira}},
  \bibinfo{author}{\bibfnamefont{R.}~\bibnamefont{Menezes}}, \bibnamefont{and}
  \bibinfo{author}{\bibfnamefont{J.}~\bibnamefont{Menezes}},
  \bibinfo{journal}{Physics Letters B} \textbf{\bibinfo{volume}{681}},
  \bibinfo{pages}{282 } (\bibinfo{year}{2009}).

\bibitem[{\citenamefont{Press et~al.}(1989)\citenamefont{Press, Ryden, and
  Spergel}}]{1989ApJ...347..590P}
\bibinfo{author}{\bibfnamefont{W.}~\bibnamefont{Press}},
  \bibinfo{author}{\bibfnamefont{B.}~\bibnamefont{Ryden}}, \bibnamefont{and}
  \bibinfo{author}{\bibfnamefont{D.}~\bibnamefont{Spergel}},
  \bibinfo{journal}{\apj} \textbf{\bibinfo{volume}{347}}, \bibinfo{pages}{590}
  (\bibinfo{year}{1989}).

\bibitem[{\citenamefont{{A. Vilenkin and E.P.S. Shellard}}(1994)}]{VS}
\bibinfo{author}{\bibnamefont{{A. Vilenkin and E.P.S. Shellard}}},
  \emph{\bibinfo{title}{{\textit{Cosmic strings and other topological
  defects}}}} (\bibinfo{publisher}{{Cambridge University Press}},
  \bibinfo{year}{1994}).

\bibitem[{\citenamefont{Garagounis and Hindmarsh}(2003)}]{PhysRevD.68.103506}
\bibinfo{author}{\bibfnamefont{T.}~\bibnamefont{Garagounis}} \bibnamefont{and}
  \bibinfo{author}{\bibfnamefont{M.}~\bibnamefont{Hindmarsh}},
  \bibinfo{journal}{Phys. Rev. D} \textbf{\bibinfo{volume}{68}},
  \bibinfo{pages}{103506} (\bibinfo{year}{2003}).

\bibitem[{\citenamefont{Avelino et~al.}(2005)\citenamefont{Avelino, Oliveira,
  and Martins}}]{Avelino20051}
\bibinfo{author}{\bibfnamefont{P.}~\bibnamefont{Avelino}},
  \bibinfo{author}{\bibfnamefont{J.}~\bibnamefont{Oliveira}}, \bibnamefont{and}
  \bibinfo{author}{\bibfnamefont{C.}~\bibnamefont{Martins}},
  \bibinfo{journal}{Physics Letters B} \textbf{\bibinfo{volume}{610}},
  \bibinfo{pages}{1 } (\bibinfo{year}{2005}).

\bibitem[{\citenamefont{Avelino et~al.}(2007)\citenamefont{Avelino, Martins,
  Menezes, Menezes, and Oliveira}}]{Avelino200763}
\bibinfo{author}{\bibfnamefont{P.}~\bibnamefont{Avelino}},
  \bibinfo{author}{\bibfnamefont{C.}~\bibnamefont{Martins}},
  \bibinfo{author}{\bibfnamefont{J.}~\bibnamefont{Menezes}},
  \bibinfo{author}{\bibfnamefont{R.}~\bibnamefont{Menezes}}, \bibnamefont{and}
  \bibinfo{author}{\bibfnamefont{J.}~\bibnamefont{Oliveira}},
  \bibinfo{journal}{Physics Letters B} \textbf{\bibinfo{volume}{647}},
  \bibinfo{pages}{63 } (\bibinfo{year}{2007}).

\bibitem[{\citenamefont{Battye and Moss}(2006)}]{PhysRevD.74.023528}
\bibinfo{author}{\bibfnamefont{R.}~\bibnamefont{Battye}} \bibnamefont{and}
  \bibinfo{author}{\bibfnamefont{A.}~\bibnamefont{Moss}},
  \bibinfo{journal}{Phys. Rev. D} \textbf{\bibinfo{volume}{74}},
  \bibinfo{pages}{023528} (\bibinfo{year}{2006}).

\bibitem[{\citenamefont{Avelino et~al.}(2008)\citenamefont{Avelino, Martins,
  Menezes, Menezes, and Oliveira}}]{PhysRevD.78.103508}
\bibinfo{author}{\bibfnamefont{P.}~\bibnamefont{Avelino}},
  \bibinfo{author}{\bibfnamefont{C.}~\bibnamefont{Martins}},
  \bibinfo{author}{\bibfnamefont{J.}~\bibnamefont{Menezes}},
  \bibinfo{author}{\bibfnamefont{R.}~\bibnamefont{Menezes}}, \bibnamefont{and}
  \bibinfo{author}{\bibfnamefont{J.}~\bibnamefont{Oliveira}},
  \bibinfo{journal}{Phys. Rev. D} \textbf{\bibinfo{volume}{78}},
  \bibinfo{pages}{103508} (\bibinfo{year}{2008}).

\end{thebibliography}

\end{document}